\documentclass[fleqn,usenatbib]{mnras}

\usepackage{newtxtext,newtxmath}
\usepackage[T1]{fontenc}
\usepackage{ae,aecompl}

\usepackage{graphicx}	
\usepackage{amsmath}	
\usepackage{gensymb}    

\usepackage{xcolor}

\title[Planes of Satellite Galaxies]{Planes of Satellite Galaxies in the {\normalfont\textsc{\textbf{Magneticum Pathfinder}}} Simulations}

\author[Pascal U. F\"orster et al.]{Pascal U. F\"orster$^{1,2}$\thanks{E-mail: pascal.foerster.19@ucl.ac.uk (PUF)},
Rhea-Silvia Remus$^{2}$,
Klaus Dolag$^{2,3}$,
Lucas C. Kimmig$^{2}$,\newauthor
Adelheid Teklu$^{2}$,
and Lucas M. Valenzuela$^{2}$
\\
$^{1}$Department of Physics and Astronomy, University College London, London WC1E 6BT, UK\\
$^{2}$Universitäts-Sternwarte, Fakultät für Physik, Ludwig-Maximilians-Universität München, Scheinerstr.1, 81679 München, Germany \\
$^{3}$Max-Planck-Institut f\"ur Astrophysik, Karl-Schwarzschild-Stra{\ss}e 1, 85741 Garching, Germany
}

\date{Accepted XXX. Received YYY; in original form ZZZ}

\pubyear{2022}

\begin{document}
\label{firstpage}
\pagerange{\pageref{firstpage}--\pageref{lastpage}}
\maketitle

\begin{abstract}
Planes of satellites are observed around many galaxies. However, these observations are still considered a point of tension for the $\Lambda$CDM paradigm. We use the fully hydrodynamical cosmological $\Lambda$CDM state-of-the-art simulation \textsc{Magneticum Pathfinder} to investigate the existence of such planes over a large range of haloes, from Milky Way to galaxy cluster masses. To this end, we develop the \textit{Momentum in Thinnest Plane} (MTP) method to identify planes and quantify the properties of their constituent satellites, considering both position and momentum. We find that thin planes ($20\%\cdot R_\mathrm{halo}$) containing at least $50\%$ of the total number of satellites can be found in almost all systems. In Milky Way mass-like systems, around 86\% of such planes are even aligned in momentum ($90\%$ of the total satellite momentum), where the fraction is smaller if more satellites are required to be inside the plane. We further find a mass dependency, with more massive systems exhibiting systematically thicker planes. This may point towards the change from continuous accretion of small objects along filaments and sheets for less massive haloes to the accretion of large objects (e.g., major mergers) dominating the growth of more massive haloes. There is no correlation between the existence of a plane and the main galaxy's morphology. Finally, we find a clear preference for the minor axes of the satellite planes and the host galaxy to be aligned, in agreement with recent observations.
\end{abstract}




\section{Introduction}
Our own galaxy, the Milky Way (MW), is surrounded by several satellite galaxies, among which the most famous ones are the Magellanic Clouds. Both satellite galaxies and also several globular clusters of the MW have been found to be distributed anisotropically in a thin plane-like structure \citep{lyndenbell:1976}, which has been dubbed the \textit{Vast Polar Structure} (VPOS; \citealp{kroupa_2005_great_disk_mw_cosmo,pawlowski_2012_vpos_mw}). Even with more satellites discovered, many of them are still found be be well aligned with the VPOS plane \citep{pawlowski_2016_alignment_of_sdss_satellites_survey_footprint,pawlowski_2018_planes_of_satellite_galaxies_problem}. Similarly, a plane of satellites has also been observed around our neighbouring galaxy Andromeda \citep{conn_ibata_2012_bayesian_approach_distances_m31,ibata_2013_vast_thin_plane_andromeda}.

More planes around galaxies beyond the local group have been observed, for example around Centaurus~A \citep{tully_2015_two_planes_centaurus_a,mueller_2016_testing_two_planes_centaurus_a}, M~101 \citep{mueller:2017_m101}, and M~83 \citep{mueller:2018_m83}.
Such planes of satellites are difficult to observe given the low-surface brightness nature of most of the satellite galaxies, and thus studies have been limited to the local Universe.
Recently, using data from the MATLAS survey \citep{duc:2015}, \citet{Heesters_2021_MATLAS} analysed 119 satellite systems and found planes in 31 of them, suggesting a probability of about 25\% for a given satellite system to contain a satellite plane. Additionally, recent studies of galaxy clusters from simulations and observations found evidence for anisotropic plane-like structures in the satellite distributions of galaxy clusters, albeit these were generally thicker than planes observed around galaxies \citep{gu:2022}.

The origin of such satellite planes is still under debate. From the hierarchical growth scenario of cosmological structure formation, assembly of galaxies onto central haloes happens along the filaments and sheets that connect them. Therefore, a certain anisotropic distribution of infall directions is an immediate consequence of the hierarchical growth scenario \citep[e.g.,][]{aubert:2004,wang:2005}. In such a scenario, the satellite planes would simply reflect the directions of infall from the connected filaments \citep{libeskind_2005_distribution_of_satellite_galaxies_pancake,shao_2018_multiplicity_anisotropy_satellite_accretion}. Alternatively, and also in agreement with the hierarchical growth scenario of the $\Lambda$CDM universe, satellite planes could originate from the infall of a group of galaxies that subsequently get disrupted due to the tidal field of the central host galaxy \citep[e.g.,][]{lynden-bell_1995_ghostly_streams_galaxy_halo_formation}, similar to the disruption of dwarf galaxies resulting in the build-up of a stream around the host.
Finally, such satellite planes could also be assembled from tidal dwarf galaxies formed during the gas-rich merger of two larger galaxies \citep[e.g.,][]{wetzstein_2007_dwarf_galaxies_in_tidal_tails}, albeit in such a scenario one would expect such satellite planes to be preferentially hosted by spheroidal galaxies. For more details on these three formation scenarios, see \citet{pawlowski_2018_planes_of_satellite_galaxies_problem}, \citet{welker_2018_caught_in_the_rhythm_1_settle_into_plane}, and \citet{Heesters_2021_MATLAS}.

While these three scenarios all give a physical reason for the appearance of satellite planes around galaxies, one other possibility is that such structures are simply coincidental alignments. Using dark matter-only simulations of 21 Andromeda mass-like haloes, \citet{buck:2016} could successfully reproduce thin planes of satellites with similar rotation properties as those observed for Andromeda; however, they also showed that they were a transient feature rather than long lived. 

Cosmological simulations so far give non-conclusive answers to the question of the existence and origin of satellite planes around galaxies. On the one hand, \citet{welker_2017_caught_in_the_rhythm_2_comp_alignment} and \citet{welker_2018_caught_in_the_rhythm_1_settle_into_plane} report that they commonly find satellite planes around galaxies with similar features as observations in the HorizonAGN hydrodynamical cosmological simulations, with a clear alignment of the galaxy and plane minor axes, similar to earlier results by \citet{dong_2014_distribution_of_satellites_around_centrals_cosmo_sims}. Further, \citet{cautun:2015} report for a set of isolated high resolution cosmological simulations of MW and Andromeda mass like haloes even more prominent satellite planes than those observed for MW and Andromeda in 10\% of their galaxies. Similarly, using the Aquarius and Millenium-II simulations, \citet{wang_2013_spatial_distribution_satellites_lcdm} find that such thin satellite planes can be reproduced in simulations for about 5--10\% of the galaxies and that thicker planes are rather common. Such planes are built up through accretion from the cosmic web, with group accretion leading to thinner planes than global accretion.

On the other hand, and using the Millenium-II simulations as well, \citet{pawlowski_2014_co-orbiting_satellite_galaxy_still_in_conflict} show that they find planes with rotational properties and as thin as those observed around the MW and Andromeda only for significantly less than 1\% of the haloes, claiming that such thin rotating planes are in conflict with the $\Lambda$CDM paradigm. This is supported by results using the Illustris and Illustris-TNG simulations \citep{mueller:2018,mueller:2021}.

In this work we study galaxies from the fully hydrodynamical cosmological simulations suite \textsc{Magneticum Pathfinder}, aiming to answer the question of how common planes of satellites are in haloes from MW mass up to galaxy cluster mass scales. Therefore, we first introduce the simulation in Sec.~\ref{sec:magneticumBox4Uhr}, followed by an analysis of the satellite mass function found for the simulation in Sec.~\ref{sec:satmassfct}. In Sec.~\ref{sec:mtp_scheme} we introduce the \textit{Momentum in Thinnest Plane} method, MTP, the method used to find satellite planes around galaxies using both position and also angular momentum information of the satellites. The results of our study, namely the connection of properties of the central galaxies to the surrounding satellite plane properties, are presented in Sec.~\ref{sec:galprop}. Finally, we summarise and conclude our results in Sec.~\ref{sec:sumconc}.

\section{The Magneticum simulations}\label{sec:magneticumBox4Uhr}
In this study we use galaxies selected from the state-of-the-art, cosmological hydrodynamical simulation suite \textsc{Magneticum Pathfinder}\footnote{\url{www.magneticum.org}}, a set of cosmological volumes simulated at different resolutions, performed with an improved developers' version \citep[see][for details on the numerical scheme]{beck_2016_improved_sph_for_cosmo} of the N-body/SPH code \textsc{Gadget-3}, which in turn is an updated version of the well-known open-source code \textsc{Gadget-2} \citep{springel_2005_gadget_2}. 
The simulations follow a standard $\Lambda$CDM cosmology adopting parameters according to the WMAP-7 cosmology \citep{2011ApJS..192...18K} ($h$, $\Omega_{M}$, $\Omega_{\Lambda}$, $\Omega_{b}$, $\sigma_{8}$) set to
($0.704$, $0.272$, $0.728$, $0.0451$, $0.809$).

\textsc{Magneticum} features a wide range of physical processes \citep[see][for details]{Hirschmann14a,teklu_2015_angular_momentum_and_galactic_dynamics,Dolag17}
which are important for studying the formation of AGN, galaxies, and galaxy clusters, with a detailed treatment of key processes that are known to control galaxy evolution.
Especially, detailed properties of galaxies of different morphologies can be recovered and studied, for example their angular momentum properties and the evolution of the stellar mass--angular momentum relation with redshift \citep{teklu_2015_angular_momentum_and_galactic_dynamics,Teklu16}, stellar kinematics of early type galaxies \citep{Schulze18,Schulze20}, the size-mass relations and their evolution \citep[e.g.,][]{Remus16,Remus2017b}, global properties like the fundamental plane \citep{Remus16} or dark matter fractions \citep{Remus2017b}, in-situ and ex-situ fractions \citep{Remus21}, the baryon conversion efficiency \citep[e.g.,][]{Steinborn15,Teklu17}, as well as chemical properties \citep{Dolag17,Kudritzki2021}.

In this study, we use the smallest but highest resolved simulation volume \textit{Box4 (uhr)}, which covers a volume of $(68\; \mathrm{Mpc})^{3}$, initially sampled with $2\cdot576^{3}$ particles (dark matter and gas) with a mass resolution of $m_\mathrm{gas} = 7.3\times10^{6}\;\mathrm{M_\odot}\;h^{-1}$ for the gas and $m_\mathrm{*} = 1.3\times10^{6}\;\mathrm{M_\odot}\;h^{-1}$ for stellar particles, with a gravitational softening of $0.7\;\mathrm{kpc}\;h^{-1}$.
Galaxies in the simulation are identified using a modified version of \textsc{subfind} \citep{springel_2001_populating_cluster_subfind,dolag_2009_substructures_in_hydro_cluster_sims}. We split galaxies into central and satellite galaxies, with \textit{central galaxy} being the main galaxy of a halo identified by \textsc{subfind}, and the \textit{satellite} galaxies of this halo being all other galaxies with stellar content identified by \textsc{subfind} in this given halo.  
Our final sample consists of $\mathcal{N}_h = 618$ central galaxies with total halo masses above $M_\mathrm{tot} = 7.1 \times 10^{11} \, \mathrm{M_{\odot}}$, with the largest halo being a small galaxy cluster of $M_\mathrm{tot} = 2.3 \times 10^{14} \, \mathrm{M_{\odot}}$.

Central galaxies are classified into disks, ellipticals, and intermediates according to their $b$-value:
\begin{equation}
    b = \log_{10} \left( \frac{j_\mathrm{*}}{\mathrm{kpc \, km \, s}^{-1}} \right) - \frac{2}{3} \log_{10} \left( \frac{M_\mathrm{*}}{\mathrm{M_\odot}} \right) \; ,
    \label{eq:bval}
\end{equation}
a parametrisation of the $M_\mathrm{*}$-$j_\mathrm{*}$ relation by \cite{teklu_2015_angular_momentum_and_galactic_dynamics}. The threshold values and numbers of each category in our sample are listed in Table~\ref{tab:b-values_in_our_data}.
\begin{table}
    \centering
    \caption{Range of $b$-values and number of systems, $N_\mathrm{sys}$, for our three $b$-value bins describing disky, intermediate and spheroidal central galaxies. See Equation~\ref{eq:bval} and \citet{teklu_2015_angular_momentum_and_galactic_dynamics} for the definition of the $b$-value. 
    }
    \begin{tabular}{lll}
        \hline
        central galaxy    & $b$-value bins           & $N_\mathrm{sys}$\\ 
        \hline
        disk              & $b > -4.375$             & 152 \\
        intermediate      & $-4.75 < b \leq -4.375$  & 216 \\
        spheroid          & $b \leq -4.75$           & 254 \\ 
        \hline
    \end{tabular}
    \label{tab:b-values_in_our_data}
\end{table}

\begin{figure*}
    \centering
    \includegraphics[width=1.8\columnwidth]{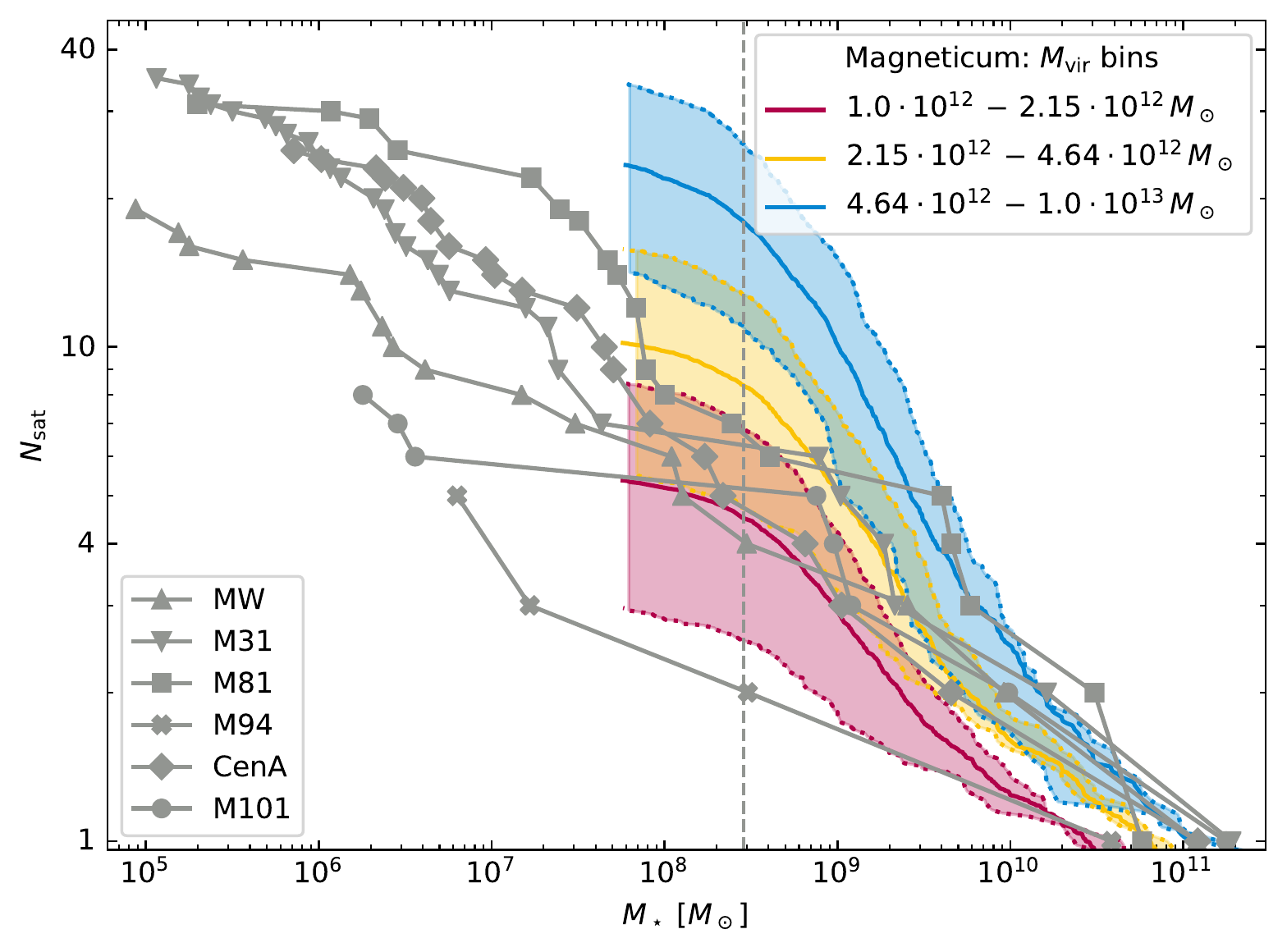}
    \caption{Satellite mass function of \textsc{Magneticum} galaxies stacked in three virial mass bins (coloured), as well as the satellite mass functions of the six nearby galaxies found by \citet{crnojevic_2019_faint_end_cen_a_luminosity_function} (grey, with different symbols). The solid coloured lines denote the median satellite number $N_\mathrm{sat}$ for the three virial mass bins in \textsc{Magneticum} as given in the legend. The correspondingly shaded regions show the respective scatter, defined by the median number of satellites for the 25\% most and least satellite-rich galaxies in each $M_\mathrm{vir}$ bin. The horizontal dashed line shows our confidence limit of at least 100 star particles of mass $2 \times 10^8 \, h^{-1} \;\mathrm{M_\odot}$ each in any given satellite.}
    \label{fig:satelliteMassFunction_comparision}
\end{figure*}

\section{Satellite mass functions}\label{sec:satmassfct}

One long standing debate with respect to the masses of satellite galaxies from simulations in comparison with those from observations has been centred around the question whether there are too many and too massive satellite galaxies found in simulations as compared to observations. Early studies \citep[e.g.,][]{moore:1999} compared dark matter-only simulations to observations and found that the simulated galaxies had many more satellite galaxies than observations of MW-type systems. By now, observations have improved and the satellite galaxy systems of galaxies in the local volume \citep[e.g.,][]{crnojevic_2019_faint_end_cen_a_luminosity_function,kim:2022} and even outside the local group environment have been measured \citep[e.g., the SAGA survey][]{mao:2021}, showing that there is a wide range of satellite mass functions, with the MW being just an average galaxy.

Since the advent of baryonic simulations, a more direct comparison between simulations and observations has become possible, effectively showing that with the inclusion of baryonic physics the discrepancies between simulations and observations vanish \citep[e.g.,][]{font:2021}, albeit the reproduction of the radial distribution of the MW satellites in various simulations is still problematic; meanwhile, the Andromeda satellite galaxy distribution is well within the predictions from simulations \citep{carlsten:2020}.

Figure~\ref{fig:satelliteMassFunction_comparision} depicts the stacked satellite mass functions from the simulated galaxies in three different halo virial mass bins, from MW-like $10^{12}\;\mathrm{M_\odot}$ to small group mass haloes of $10^{13}\;\mathrm{M_\odot}$ with the bin size selected such that each bin contains the same number of central galaxies. The solid lines show the median satellite mass functions found for the galaxies in each halo mass bin, and the shaded areas mark the scatter of satellite mass functions. For comparison, observations of the satellite systems of five galaxies in addition to the MW galaxy, as compiled by \citet{crnojevic_2019_faint_end_cen_a_luminosity_function}, are included as grey lines and symbols.

As can clearly be seen, the scatter in the lowest halo mass bin is the largest for the lower-mass end of the satellite mass function, and nearly reaches the low satellite numbers observed for the very satellite galaxy poor system M~94 \citep{smercina:2018}, and well covering the distribution found for the MW. Even though the resolution of our simulation is not high enough to cover the satellite galaxy population below $10^8\;\mathrm{M_\odot}$, the fact that we can successfully reproduce the observed ranges of satellite mass functions is important as we will now use these satellite galaxies to investigate the occurrence and properties of satellite galaxy planes in the simulations.

\section{Finding Satellite Planes: Momentum in Thinnest Plane scheme}\label{sec:mtp_scheme}
In this section, we direct our focus to the task of finding and implementing a suitable scheme to identify and analyse planes of satellite galaxies around their central galaxies. To this end, we have developed the \textit{Momentum in Thinnest Plane} (MTP) method, which determines and classifies satellite structures by the momentum alignment of satellites to their position-wise best fitted, thinnest planes.

The Momentum in Thinnest Plane (MTP) scheme is applied to every system in our ensemble of $\mathcal{N}_h = 618$ haloes individually. Generally, a plane of satellites is defined by their similar regions in phase space. This means there are two determining factors: whether the satellite positions are constrained in space within a plane, and whether their momenta align with the orientation of the plane. Therefore the MTP scheme consists of two steps. First, for a halo with $N$ satellites, the thinnest $N-1$ planes which contain $n=[2,\, N]$ satellites are determined, always including the central galaxy. There is then one thinnest plane containing $n=2$, one with $n=3$ and so on up to a plane containing all $N$ satellites. Second, for these $N-1$ planes a measure is developed to quantify how strongly the movements of the constituent satellites lie within the plane. We will describe both steps in more detail below, with the second step starting from Sec.~\ref{subsec:computing_in-plane_momentum}.

\subsection{Step 1: Distances from a Plane}\label{subsec:distances_from_plane}

\begin{figure*}
    \centering
    \includegraphics[width=1.5\columnwidth]{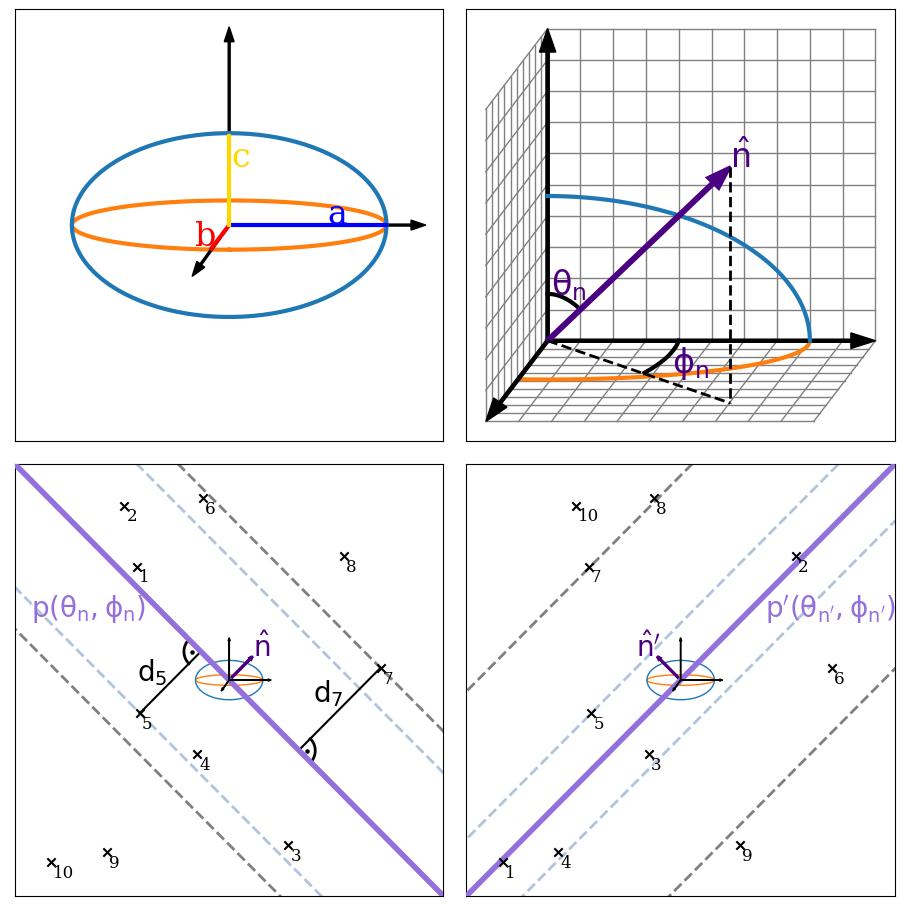}
    \caption{Overview of the method for determining the thinnest plane for a given number of satellites which it should contain. \textit{Top left:} The chosen coordinate system as defined by the central galaxy ellipsoid. $X$, $Y$ and $Z$-axes align with the major, middle and minor axes respectively (though note that here $b$ represents the \textit{negative} $Y$-axis to better show the $X$ and $Z$-axes). \textit{Top right:} The normal vector $\mathrm{\hat{n}}$ as defined via the two spherical coordinates $\mathrm{\theta_n}$ and $\mathrm{\phi_n}$, representing the angle of $\mathrm{\hat{n}}$ to the $Z$ and $X$-axes. \textit{Bottom left:} Zoomed-out look of the central galaxy and its satellites (black crosses). For any plane $p(\mathrm{\theta_n},\mathrm{\phi_n})$ (purple solid line) the distances of the satellites $i$ to plane $p$ are given by the orthogonal distances $d_\mathrm{i}$. Note that here $i$ are already sorted by their distances, such that $i=1$ is the closest satellite to the plane. The thickness of the plane containing $n=5$ satellites is then given by Equation~\ref{eq:methods.dpn} as $D_{p,5}=2\cdot d_\mathrm{5}$ (light-blue dashed lines), while for $n=7$ the thickness is $D_{p,7}=2\cdot d_\mathrm{7}$ (grey dashed lines). \textit{Bottom right:} The same as \textit{bottom left} for a different plane $p'(\mathrm{\theta_{n'}},\mathrm{\phi_{n'}})$. Here $D_{p',5}$ (light-blue dashed lines) is much thinner than for $p$, while conversely $D_{p',7}$ (grey dashed lines) is relatively larger.}
    \label{fig:MTP_schempos}
\end{figure*}

To start with, a coordinate system must be defined. We choose the coordinate system given by the central galaxy as shown in the top left panel of Figure~\ref{fig:MTP_schempos}, with the angular momentum of the central galaxy pointing in the direction of the $x$-axis. In general, an arbitrarily oriented plane $p$ is defined by its normal vector $\mathrm{\hat{n}}$ and the distance to the origin of the coordinate system. As the planes of satellites should be defined relative to their central galaxy, we only consider planes crossing the origin; and as such, our planes are entirely defined by the normal vector $\mathrm{\hat{n}}$.

Given the coordinate system, it is useful to transform into spherical coordinates. A vector is then defined by three spherical coordinates as $\Vec{r}=(r,\theta,\phi)$, where $\theta$ and $\phi$ are the angles between $\Vec{r}$ and the $Z$ and $X$-axes, so that
\begin{equation}
    \theta\equiv    \arccos\left(\frac{\Vec{r}\cdot\mathrm{\hat{z}}}{|\Vec{r}|}\right), \ \ \phi\equiv \arccos\left(\frac{\Vec{r}\cdot\mathrm{\hat{x}}}{|\Vec{r}|}\right), \label{eq:methods.angles}
\end{equation}
while $r=|\Vec{r}|$ is its length. As any normal vector fulfils $|\mathrm{\hat{n}}|=1$ by definition, a unique plane is fully characterised via the two spherical coordinates of its polar angle $\mathrm{\theta_n}$ and azimuthal angle $\mathrm{\phi_n}$; therefore $p=p(\mathrm{\theta_n},\mathrm{\phi_n})$. This is shown in the top right panel of Figure~\ref{fig:MTP_schempos}.

The goal is then to determine the thinnest possible plane $p_{\min(n)}$ containing a given number of satellites $n$. Thus, the distances from the satellites to any considered plane $p$ need to be determined. For a satellite $i$ at position $\vec{r}_\mathrm{i}$, the shortest such distance is given by the orthogonal distance as
\begin{equation}
    d_\mathrm{i}\equiv\vec{r}_\mathrm{i}\cdot\mathrm{\hat{n}}, \label{eq:methods.distance}
\end{equation}
which represents the line going away perpendicularly from the plane directly toward the satellite, parallel to the normal vector. This is illustrated in the bottom left panel of Figure~\ref{fig:MTP_schempos} for an example plane $p(\mathrm{\theta_n},\mathrm{\phi_n})$ (purple solid line) with $N=10$ satellites (black crosses). The shortest distances of satellites 5 and 7 are given as the solid black lines with length $d_\mathrm{5}$ and $d_\mathrm{7}$, respectively.

If the satellite position is given by the Cartesian coordinates as $\vec{r}_\mathrm{i}=(x_\mathrm{i},y_\mathrm{i},z_\mathrm{i})$, then substituting the polar coordinates 
\begin{equation}
    n_\mathrm{x}=\sin\theta_\mathrm{n}\cdot\cos\phi_\mathrm{n}\,, \ \ n_\mathrm{y}=\sin\theta_\mathrm{n}\cdot\sin\phi_\mathrm{n}\,, \ \ n_\mathrm{z}=\cos\theta_\mathrm{n}, \label{eq:methods.polar}
\end{equation}
into Equation~\ref{eq:methods.distance} gives:
\begin{equation}
    d_\mathrm{i}=x_\mathrm{i}\cdot \sin\theta_\mathrm{n}\cos\phi_\mathrm{n} + y_\mathrm{i}\cdot \sin\theta_\mathrm{n}\sin\phi_\mathrm{n} + z_\mathrm{i}\cdot \cos\theta_\mathrm{n}. \label{eq:methods.distance_polar}
\end{equation}

Note that in the following the subscripts of the angles are dropped, as $(\theta,\phi)$ will always be defined relative to the normal vector $\mathrm{\hat{n}}$.

\subsubsection{Projecting spherical coordinates onto a grid}
\label{subsec:methods_projecting_spherical_coordinates_onto_grid}

Let us now consider how to project the possible planes onto a sufficiently fine grid for further analysis. Each of the above planes $p \, (\theta, \phi)$ is characterized by a set of two polar angles, $\theta$ and $\phi$. We will search for the thinnest $n$-satellite plane after putting all possible values of 
\begin{equation*}
    \theta = \left[0, \, \frac{\pi}{2} \right] \quad \mathrm{and} \quad \phi = [0, 2\pi]
\end{equation*}
onto a grid with a minimum accuracy $a$. We define $a$ such that $f = \frac{1}{a}$ is a fraction of the total radius $R$ of the system; therefore we segment the $\pi/2$-length chord of the unit circle along the range of $\theta = [0, \pi/2]$ into a minimum of 
\begin{equation*}
    \kappa = \frac{\pi/2}{f} = \frac{\pi}{2} \cdot a
\end{equation*}
bins. Since we require an integer number of equally spaced bins, we use 
\begin{equation}
    k_\theta = \lceil \kappa \rceil + 1 \quad \mathrm{and} \quad k_\phi = 4 \, \lceil \kappa \rceil \quad ,
    \label{eq:methods.number_of_theta_phi_bins}
\end{equation}
for the number of bins along $\theta$ and $\phi$.

A choice of very low accuracy of for example $a=3$ results in $k_\theta = 6$ bins along the polar angle and $k_\phi = 20$ bins along the azimuthal angle, for a total of $N_p = 120$ planes. However, this would not sample the space sufficiently, and potentially much thinner or better aligned planes could be missed. Instead, throughout this work $a=180$ is chosen, corresponding to an angular resolution of at least $0.5^\circ$. The space of spherical coordinates is thus sampled with a total of $N_p = 284 \cdot 1132 \approx 3 \times 10^5$ cells for each halo.

\subsubsection{Finding the thinnest plane}
\label{subsec:finding_thinnest_plane}

Now let us consider a halo with $N$ satellites and a central galaxy. We want to fit a plane containing the central galaxy to an increasing number $n$ out of the $N$ total number of satellites, starting with $n=2$, as the central and just one satellite do not form a well defined plane. There will always be a perfectly fitted, zero-thickness plane containing the central galaxy and \textit{any} $n = 2$ satellites, so which of the $N-1$ possible ``thinnest'' planes is selected may depend on chance and numerics in this case, and is thus of no further consideration here.

From $n=2$ we move upward in steps of one in number of satellites and, for each number $n=3, 4, \dots N$, look for the plane $p_n \, (\theta, \phi)$ through the central that features $n$ out of the total number of satellites $N$ with the smallest maximum distance from $p_n$. This is illustrated in the bottom two panels of Figure~\ref{fig:MTP_schempos} for a main halo with $N=10$ satellites. Let us assume we want the thinnest plane containing half of the satellites, so $n=5$. By varying through the spherical coordinates as described in Sec.~\ref{subsec:methods_projecting_spherical_coordinates_onto_grid} we get many potential candidate planes $p$. The bottom left depicts one such plane, where the indices of the satellites $i$ are sorted such that $i=1$ is the satellite closest to the plane. This ordering thus \textit{changes} based on the given plane -- see the different ordering of the satellites in the bottom right. By definition the thickness of the plane containing $n$ satellites is then given as:
\begin{equation}
    D_\mathrm{n,p} \equiv 2\cdot d_\mathrm{n} \label{eq:methods.dpn}
\end{equation}
where $d_\mathrm{n}$ is the distance of the $n$-th closest satellite to the given plane $p$ as per Equation~\ref{eq:methods.distance_polar}. 

For our example in the bottom two panels of Figure~\ref{fig:MTP_schempos}, the plane containing five satellites is thicker for $p$ on the left-hand side than it is for $p'$ on the right-hand side, so $D_\mathrm{5,p}>D_\mathrm{5,p'}$. This does not mean that $p'$ is always the ``thinner'' plane -- indeed it can be seen that for $n=7$ satellites, the right-hand example is preferred with $D_\mathrm{7,p}<D_\mathrm{7,p'}$! 

Consequently, when applying the above method to all $\mathcal{N}_h=618$ haloes in our ensemble, for all the $N_p$ possible planes $p$ on our $\theta$-$\phi$ grid, the thicknesses $D_\mathrm{n,p}$ of potential disks of satellites is determined for every feasible number of satellites $n$ individually. The thinnest plane of a halo for a given $n$ is then taken from the $\theta$-$\phi$ grid as the angles which produce the smallest $D_\mathrm{n,p}$, and is henceforth called $p_n$.

\begin{figure*}
    \centering
    \includegraphics[width=1.025\columnwidth]{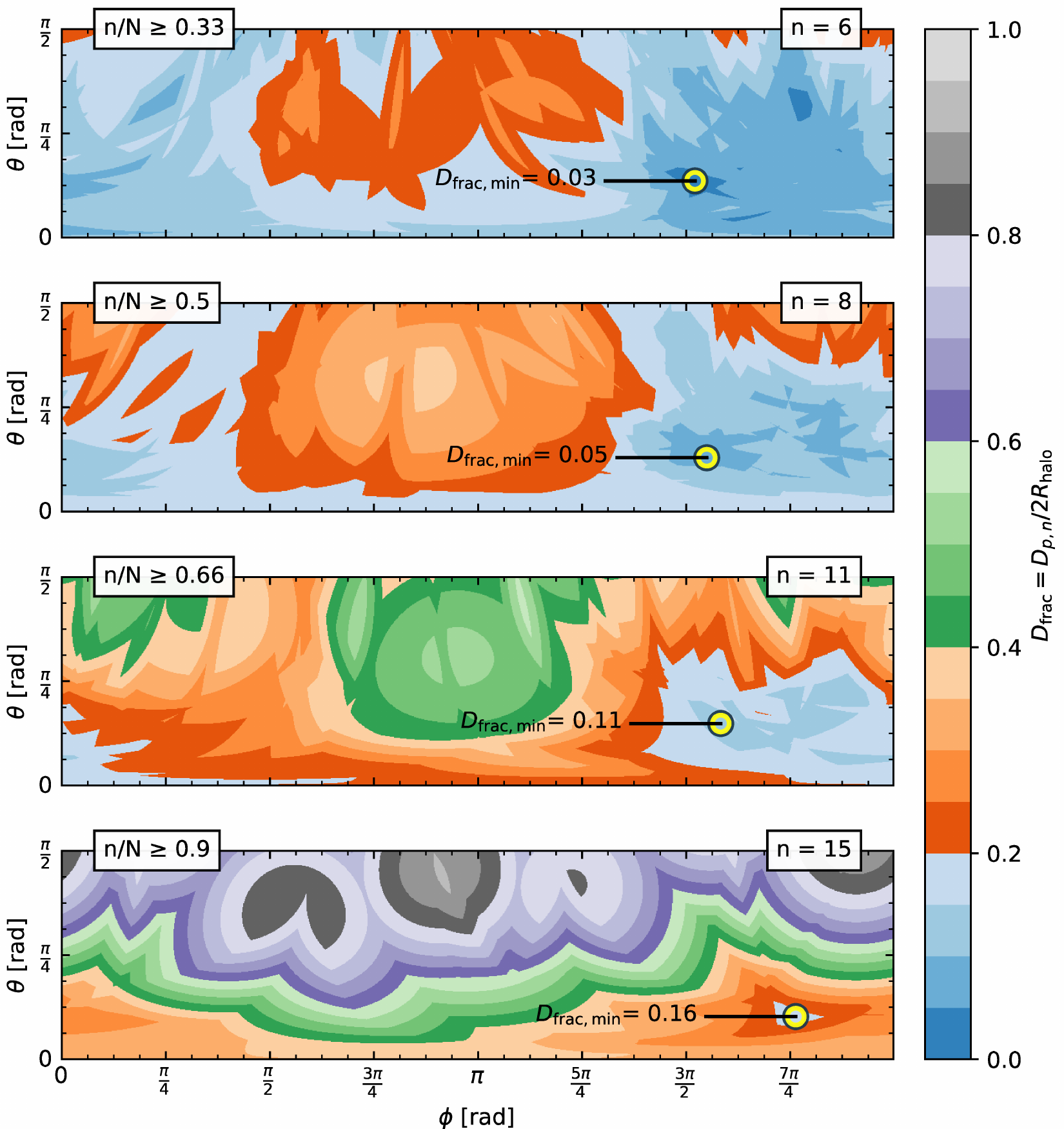}
    \quad
    \includegraphics[width=1.025\columnwidth]{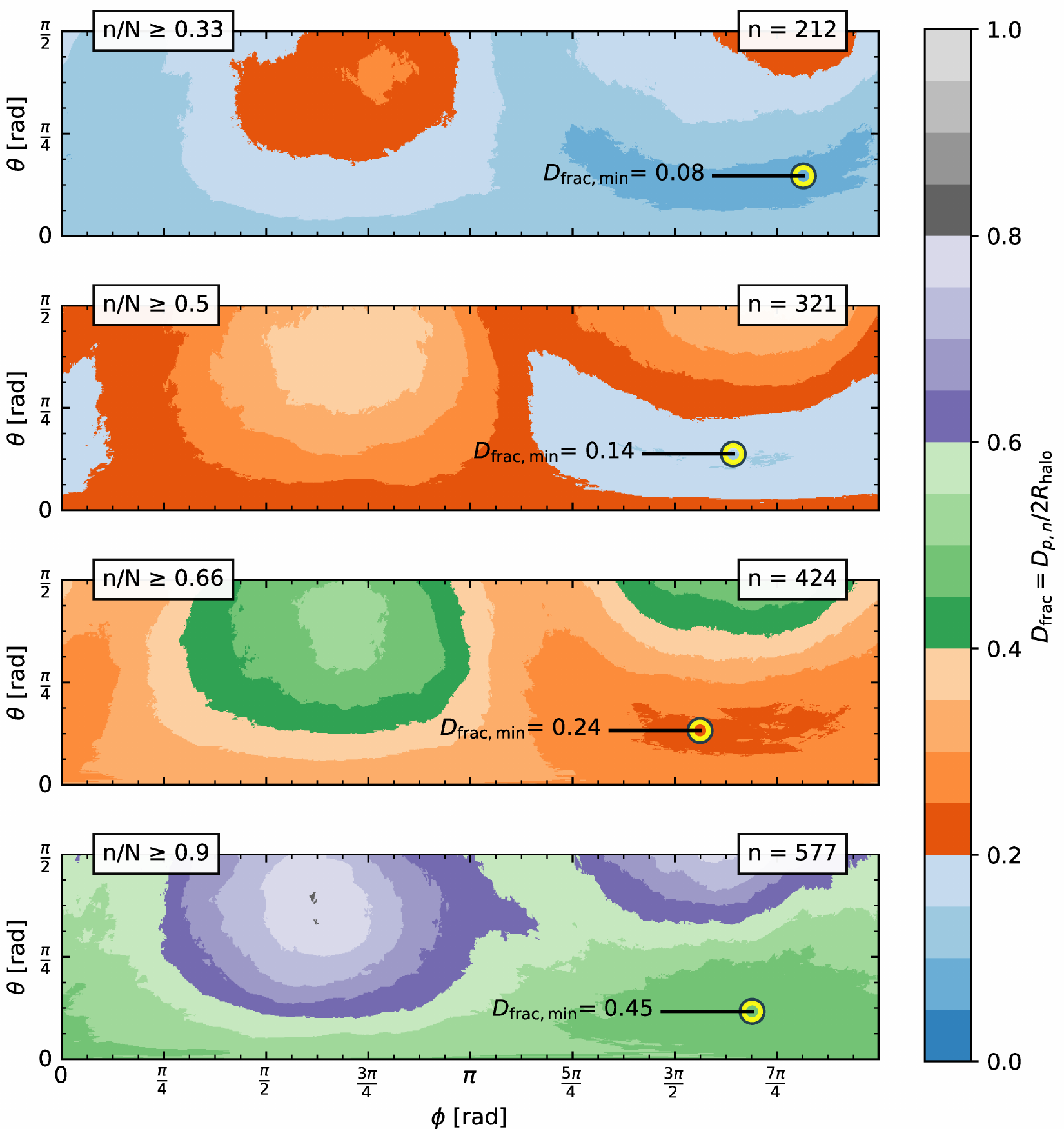}
    \caption{Maps of the thickness fractions $D_\mathrm{frac}$, containing $n/N \geq 33\%, \, 50\%, \, 66\%, \, 90\%$ of all satellites in the halo, respectively, for two example systems. The actual numbers of satellites used for the map are denoted in the upper right panel of each figure.
    The angle $\theta$ is shown against the angle $\phi$. The colours denote different plane thickness values, from thin planes in blue to thick planes (that actually would not be considered planes anymore) in grey.
    \textit{Left panels:} for an example galaxy with a small satellite system.
    \textit{Right panels:} for an example galaxy with a large satellite system.
    }
    \label{fig:thickness_maps_252}
\end{figure*}

\subsubsection{Thickness maps for two example systems} \label{subsec:results_individual_thickness_maps}

To illustrate the process of finding the position-wise thinnest plane, two example systems are shown in Figure~\ref{fig:thickness_maps_252}. The left-hand panels are for a lower mass halo of $M_\mathrm{vir} = 2.04 \times 10^{12} \;\mathrm{M_\odot}$, with a total amount of $N=16$ satellite galaxies, while the right-hand panels depict an example for a high mass, satellite rich galaxy, with a halo mass of $M_\mathrm{vir} = 9.70 \times 10^{13} \;\mathrm{M_\odot}$ and a total of $N=641$ satellite galaxies. 

Each plot of Figure~\ref{fig:thickness_maps_252} shows the grid of spherical coordinates $\theta$ ($y$-axis) and $\phi$ ($x$-axis), with the colours representing the thickness $D_\mathrm{n,p}$ normalised by the size of the halo such that the thickness fraction is
\begin{equation}
    D_\mathrm{frac}\equiv\frac{D_\mathrm{p,n}}{2\cdot R_\mathrm{halo}} \;. \label{eq:methods.dfrac}
\end{equation}
Shades of blue denote a thickness fraction of up to $20\%$, shades of red of up to $40\%$, shades of green of up to $60\%$, shades of violet of up to $80\%$, and shades of grey denote a thickness fraction of up to $100\%$.

Going from top to bottom in the columns, we require an increasing number of satellites contained within the planes, from $33\%$ over $50\%$ and $66\%$ to $90\%$ of all satellites. Generally, requiring a higher fraction of satellites $n/N$ will result in a thicker disk, which is found for all planes from the $\phi$-$\theta$ grid. Regions which in the top panel are red (i.e., a thickness of up to $40\%\cdot R_\mathrm{halo}$) become green or purple when increasing the number of participating satellites, which is the case for both example haloes in Figure~\ref{fig:thickness_maps_252}. These thickness maps can be pictured as hills and valleys, with the thinnest plane being the lowest point on the map.

For the lower-mass halo (left-hand panels), the overall thinnest planes for each of the four $n$ have a thickness of $D_\mathrm{frac,min}=0.03, 0.05, 0.11$, and $0.16$ respectively, as indicated by the white circles. This means that even when requiring $90\%$ of all satellites to be within the plane, we find one for the lower mass halo which has a thickness of just $D_\mathrm{15,p}=0.16\cdot R_\mathrm{halo}$. This is not the case for the more massive example halo shown in the right-hand panels of Figure~\ref{fig:thickness_maps_252}, where the four $n$ result in thinnest planes of thickness $D_\mathrm{frac,min}=0.08, 0.14, 0.24$ and $0.45$. In particular,  requiring $n\geq0.9\cdot N$ results in a large jump in thickness to nearly half of $R_\mathrm{halo}$ and thus not resembling a plane anymore. This is the best-case plane, with other orientations resulting in a thickness of nearly the entire extent of the halo. Note that for the high-mass halo, even the lowest required number fraction $n/N \geq 0.5$ results in a thin plane of 212 satellites.

An interesting feature of the thickness maps is that there appear to be smooth transitions between regions on the $\phi$-$\theta$ grid, and in particular the similarities between maps when requiring higher fractions of satellites. The orientation of the thinnest plane shifts only slightly with higher fractions of satellites.

Moving forward, we select the $\theta$ and $\phi$ pair representing the global minimum in thickness for any given number of satellites $n=2, \dots N$ as the orientation of the thinnest plane $p_\mathrm{n}$ of $n$ satellites for each of our $\mathcal{N}_h=618$ haloes.

\subsection{Step 2: Computing the in-plane Momentum}\label{subsec:computing_in-plane_momentum}
Using the above method identifies the thinnest planes of satellites. However, the members of a plane of satellites should not only lie close to each other in position space but should also \textit{stay} close, that is to say their motion should be largely within the plane. 

\begin{figure*}
    \centering
    \includegraphics[width=1.7\columnwidth]{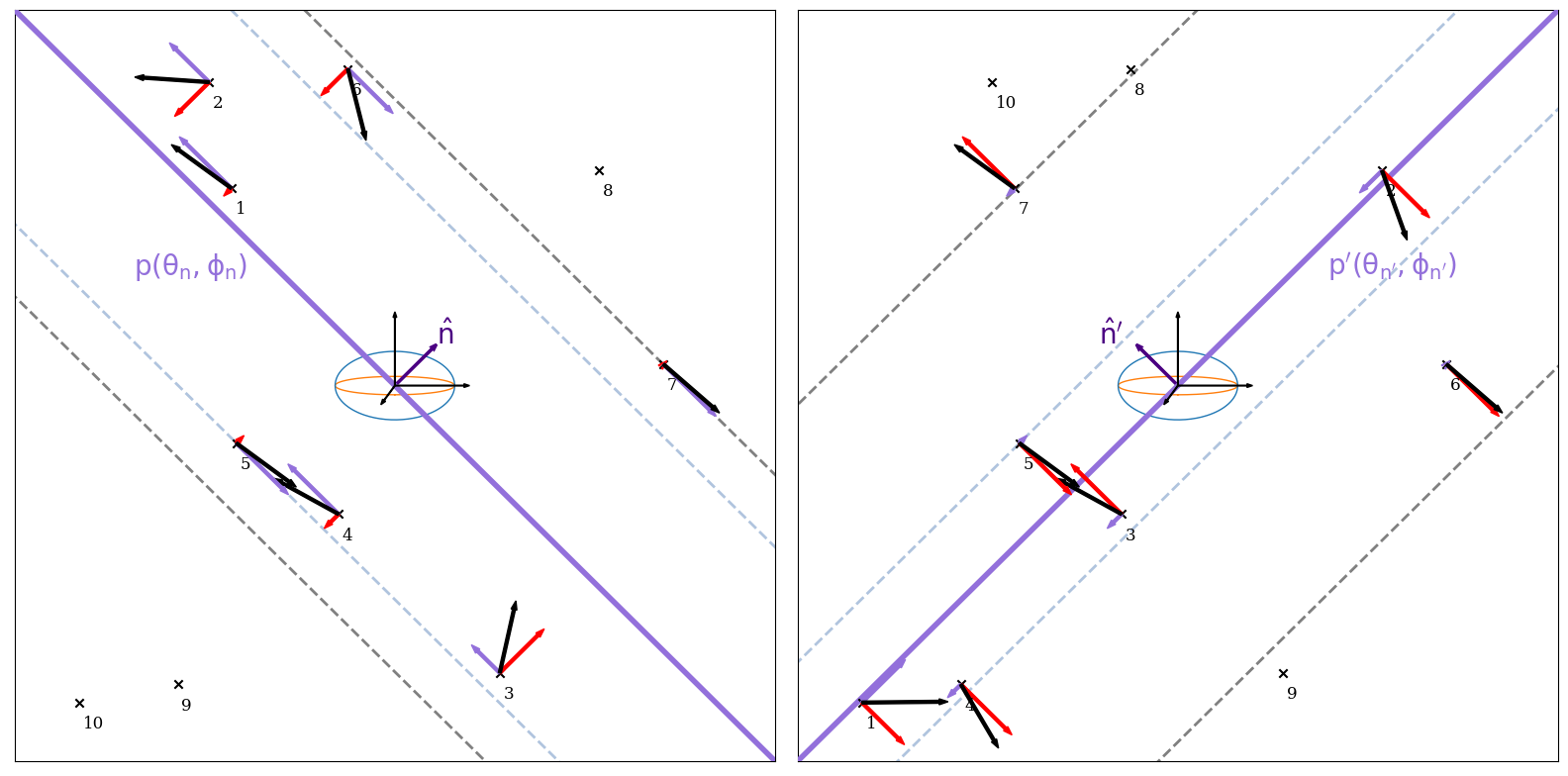}
    \caption{Zoomed-out look of the central galaxy and its satellites (black crosses), with two different planes $p$ (\textit{left}) and $p'$ (\textit{right}) as solid purple lines and the satellite indices $i$ ordered by their distances to the respective planes. The planes are the same as in the bottom two panels of Figure~\ref{fig:MTP_schempos}, with the light-blue and grey lines representing the thickness of the planes containing $n=5$ and $n=7$ satellites, respectively. For each satellite within the $n=7$ plane, its velocity vector is depicted in black and split into the components parallel (purple arrow) and perpendicular (red arrow) to the plane.}
    \label{fig:MTP_schemvel}
\end{figure*}

To illustrate this, Figure~\ref{fig:MTP_schemvel} shows the same planes $(p,p')$ as in the bottom panels of Figure~\ref{fig:MTP_schempos}. Additionally, for any satellite that lies within the $n=7$ plane, i.e., is one of the seven closest satellites to the respective planes as given by their position, the velocity vectors are depicted in black. Note that going purely by positional arguments, the right-hand side plane $p'$ for $n=5$ satellites is thinner than $p$ on the left, so $D_\mathrm{5,p}>D_\mathrm{5,p'}$. However, as can be seen now from the velocity vectors, the fraction of the velocity pointing in parallel to the plane (purple arrows) is much smaller than the perpendicular portion (red arrows) for $p'$. This means that given some time, this thinnest plane may very well dissolve. Conversely, the much thicker $n=5$ plane for $p$ in the left-hand side panel reveals that most of the velocity of its satellites \textit{does} point in parallel to the plane. In our example, this behaviour persists also for $n=7$. It could then very well be argued that $p$ potentially is a more likely real plane than $p'$ is, and we should choose $p_5=p'$. This necessitates the inclusion of the velocities when considering whether a configuration is a viable plane.

To this end, we apply our method outlined in Sect.~\ref{subsec:distances_from_plane} also to the velocity, and thereby momentum, of satellites. Again, we take a plane $p$ through the origin of the coordinate system. The velocity $\Vec{v}$ of a satellite galaxy completely moving in direction of this plane would be parallel to said plane $p$, and once shifted to the origin, geometrically lie within this plane fulfilling:
\begin{equation*}
    \Vec{v}_\mathrm{in-plane} \cdot \mathrm{\hat{n}} = 0.
\end{equation*}

Any arbitrary satellite $S_i$ that is not moving completely within the plane $p$ will have a velocity vector $\Vec{v}_i$ pointing to some degree out of said plane, and the distance from this plane in $v$-space is equivalent to the absolute value of the velocity component perpendicular to the plane, $\Vec{v}_{i\perp}$. Then the equivalent to Equation~\ref{eq:methods.distance} for velocities is
\begin{equation*}
    \Tilde{d}_i = \Vec{v}_i \cdot \mathrm{\hat{n}} = v_{i,x} \, n_x + v_{i,y} \, n_y + v_{i,z} \, n_z = |\Vec{v}_{i\perp}| \; .
\end{equation*}
For consistency, we substitute $n_x, n_y, n_z$ according to Equation~\ref{eq:methods.polar} and are left with the magnitude of motion of the satellite perpendicular to the plane,
\begin{equation}
    |\Vec{v}_{i\perp}| = v_{i,x} \cdot \cos{\theta} \, \cos{\phi} \; + \; v_{i,y} \cdot \cos{\theta} \, \sin{\phi} \; + \; v_{i,z} \cdot \sin{\theta} \: .
    \label{eq:methods.hesse_normal_perpendicular_velocity_explicit_angles}
\end{equation}
This gives us a way to test if a satellite's movement will likely carry it out of the plane $p$ previously fitted according to its and its $n-1$ companion satellites' positions from the current plane selection. 

We now use the above to characterize the agreement of motion of the $n$ satellites with the previously fitted position plane $p_n$. We first derive the absolute amount of velocity in direction of the plane $|\Vec{v}_{i\parallel}|$ from the linear combination
\begin{equation*}
    \Vec{v}_{i} = \Vec{v}_{i\parallel} + \Vec{v}_{i\perp}   \: ,
\end{equation*}
and find the fraction of velocity of satellite $S_i$ in direction of the plane to be
\begin{equation}
    \Tilde{F}_i \equiv \left|\frac{\Vec{v}_{i\parallel}}{\Vec{v}_{i}}\right| = \sqrt{1 - \left|\frac{\Vec{v}_{i\perp}}{\Vec{v}_{i}}\right|^2} \; .
    \label{eq:methods.in-plane_momentum_fraction_individual}
\end{equation}

The fraction $\Tilde{F}_i$ only applies to one satellite. This can be easily expanded to combine all $n$ satellites in the plane by means of averaging, leading to
\begin{equation*}
    \Tilde{F}_{\{n\}} \equiv \frac{1}{n} \sum^n_{i=1} \Tilde{F}_i \; .
\end{equation*}
The closer $\Tilde{F}_{\{n\}}$ is to $1$, the better the motion of the ensemble of satellites is aligned with the plane. This straightforward approach to a cumulative fraction has the disadvantage that all satellites of the set $\{S_i\}$ are weighted equally, giving the lightest satellites the same influence on our measure as the most massive ones. We remedy this by weighing the satellites by their mass, which effectively makes our new measure $F_{\{n\}}$ the in-plane momentum fraction:
\begin{equation}
    F_{\{n\}} \equiv \left({\sum^n_{i=1} m_i \, |\Vec{v}_{i\parallel}|}\right) \Big/ \left({\sum^n_{j=1} m_j \, |\Vec{v}_{j}|}\right)
    \label{eq:methods.in-plane_momentum_fraction}
\end{equation}

In a final step, we now do not only compute the in-plane momentum fraction $F_{\{n\}}$ for all $n$ satellites of our fitted $n$-satellite planes $p_n$, but also for their subsets with $\eta = 1, \dots, n$ satellites, selected in decreasing order of their individual fractions $\Tilde{F}_i$. This gives us a measure for how many of the $n$ satellites to which the plane $p_n$ was fitted position-wise agree with it momentum- and thus motion-wise. 

As a final note on the momenta here, consider a satellite with a velocity pointing to equal parts perpendicular and parallel to the plane, $\Vec{v}_{i\parallel} = \Vec{v}_{i\perp}$. From Equation~\ref{eq:methods.in-plane_momentum_fraction_individual} it follows that
\begin{equation*}
    \Tilde{F}_i=\left|\frac{\Vec{v}_{i\parallel}}{\sqrt{2}\cdot\Vec{v}_{i\parallel}}\right|=\frac{1}{\sqrt{2}}.
\end{equation*}
For a subset of satellites $\eta$ this implies that $F_{\{\eta\}} = \frac{1}{\sqrt{2}} \approx 0.7$ is a boundary above which the motions of the satellites are \textit{more} aligned with the plane than not.

\begin{figure*}
    \centering
    \includegraphics[width=2\columnwidth]{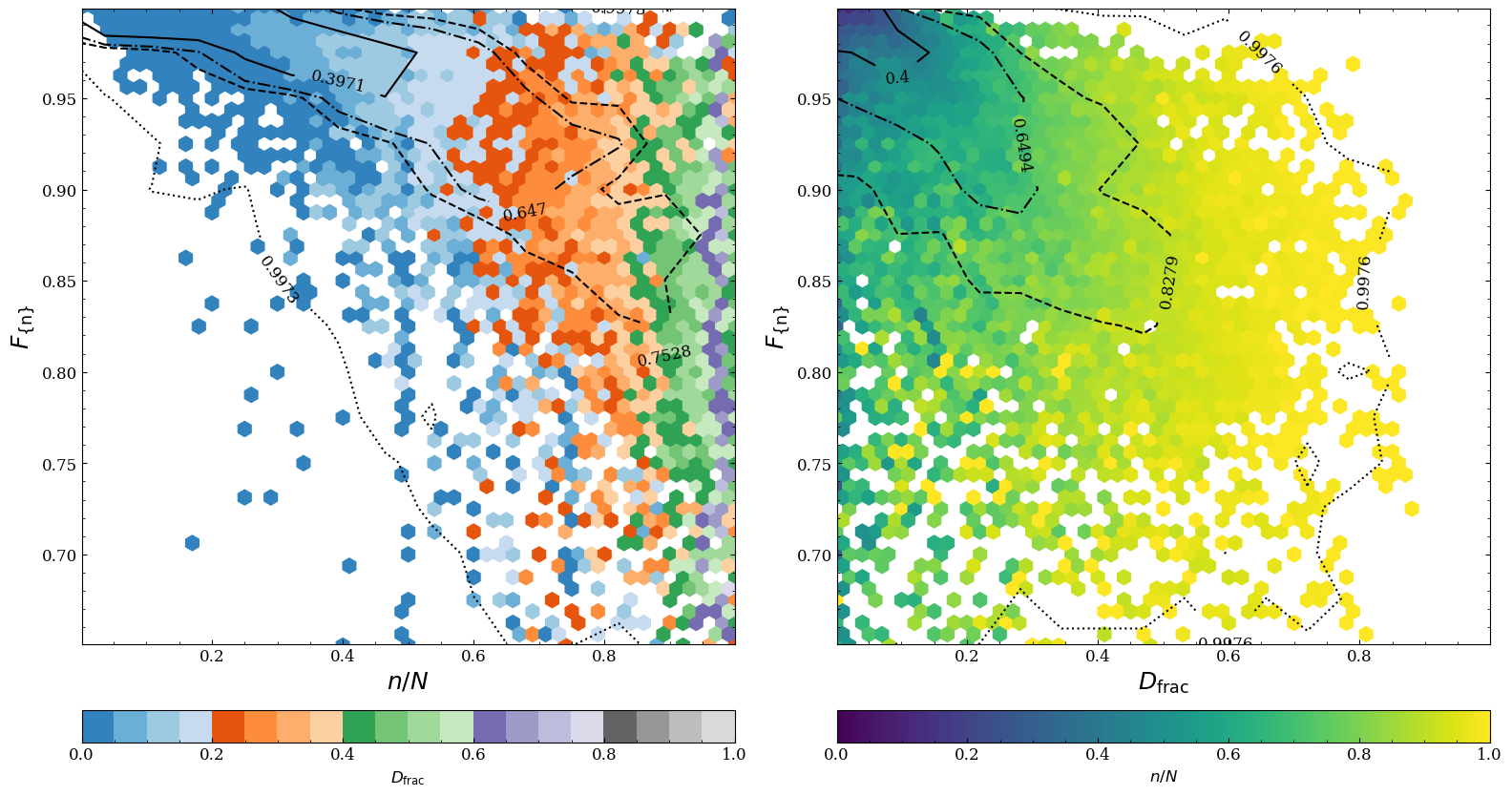}
    \caption{The in-plane momentum fraction $F_{\{n\}}$ of the thinnest planes $p_n$ as a function of satellite fraction $n/N$ coloured by the relative thickness $D_\mathrm{frac}$ (\textit{left}), and as a function of $D_\mathrm{frac}$ coloured by $n/N$ (\textit{right}). Black lines represent contours from higher (solid) to lower  density regions (dotted), with the numbers denoting the fraction of total planes contained within the contour. The plots are zoomed in on $F_{\{n\}}>0.65$, which represents $96.1\%$ of all planes $p_n$, to better show the resulting dependencies.}
    \label{fig:MTP_Endpoints}
\end{figure*}

\subsection{Cumulative Momentum in Thinnest Plane Results}
\label{subsec:cumulative_mtp_results}

In this section, we will present the combined evaluation of the momentum in thinnest plane (MTP) results. Each halo with $N$ satellites has then $N-1$ thinnest planes, containing $n=2\, \dots N$ satellites. Thus, if halo $j$ has $N_\mathrm{j}$ satellites, we obtain
\begin{equation}
    \sum_\mathrm{j=1}^{\mathcal{N}_h}(N_\mathrm{j}-1) \approx 18000
\end{equation}
data points for our galaxy sample, and each of them consists of three distinct pieces of information: the number of satellites $n$ in the position-wise fitted plane $p_n$, the thickness of that plane $D_{p,n}$, and the in-plane momentum fraction $F_{\{n\}}$ of all $n$ satellites in the plane. The number of satellites is normalised by the total satellite number $N$, and the thickness by the size of the halo as prescribed in Equation~\ref{eq:methods.dfrac}.

Figure~\ref{fig:MTP_Endpoints} shows the resulting relations, with the left panel depicting the in-plane momentum fraction $F_{\{n\}}$ as a function of satellite fraction $n/N$. The thinnest planes have been binned into hexagonal bins with the colour denoting the mean relative thickness $D_\mathrm{frac}$ of the bin. As can be seen, the distribution forms a triangle fit into the upper right corner showing that a low satellite fraction $n/N$ results in a very high amount of momentum along the plane with little scatter to lower $F_{\{n\}}$. The colour further shows that in these cases the planes are typically very thin, with $D_\mathrm{frac}<0.2$. 

When moving towards the right of the figure, i.e., to higher ratios of $n/N$, the resulting planes both scatter to lower in-plane momentum fractions as well as become noticeably thicker. The colour bands are nearly vertical -- so similar values of $D_\mathrm{frac}$ scatter up and down in $F_{\{n\}}$ -- indicating that the spread in $F_{\{n\}}$ is largely due to the higher required number of satellites as opposed to thicker planes. This can further be seen from the fact that there are planes with very high thickness (purple) that still have very high alignment of the satellite momentum with the plane, with $F_{\{n\}}>0.9$ possible.

The black contours show the fraction of planes contained within them and range from the highest density region (solid) to one containing practically all planes (dotted), with the numbers denoting the contained fraction of planes. It follows that while there is some scatter to low $F_{\{n\}}$, in actuality $75\%$ lie within a thin region curving slightly downward with increasing $n/N$ (dashed black line). Note that this downward curve means that the region of high $F_{\{n\}}$ with high $n/N$, while present, is sparsely populated with planes. There are thus few planes containing nearly all satellites while also being nearly perfectly aligned in their motions with the plane. Nonetheless, $75\%$ of all planes are found to have values of $F_{\{n\}}>0.85$. Remarkably, it follows that selecting the planes purely by being the position-wise thinnest to contain $n$ satellites results in satellite motions which are strongly aligned with the plane. More specifically, it means that these position-wise fitted planes are not a coincidental fit of satellites moving in random directions that would not be visible at a different point in time.

The right panel in Figure~\ref{fig:MTP_Endpoints} then shows the in-plane momentum fraction $F_{\{n\}}$ as a function of relative thickness $D_\mathrm{frac}$, with the colour now displaying the mean satellite fraction $n/N$ of the hexagonal bins. As hinted at before, the values of $F_{\{n\}}$ scatter largely independent of $D_\mathrm{frac}$. Instead, there is a colour trend from high satellite fractions $n/N$ (purple) to low (yellow) both when going from left to right as well as when going up to down. This means that both lower $F_{\{n\}}$ or $D_\mathrm{frac}$ will result in higher $n/N$. However, as can be seen from the contours, most of the planes are contained in the upper left region, so feature high in-plane momentum fractions in thin planes.

\begin{figure*}
    \centering
    \includegraphics[width=\textwidth]{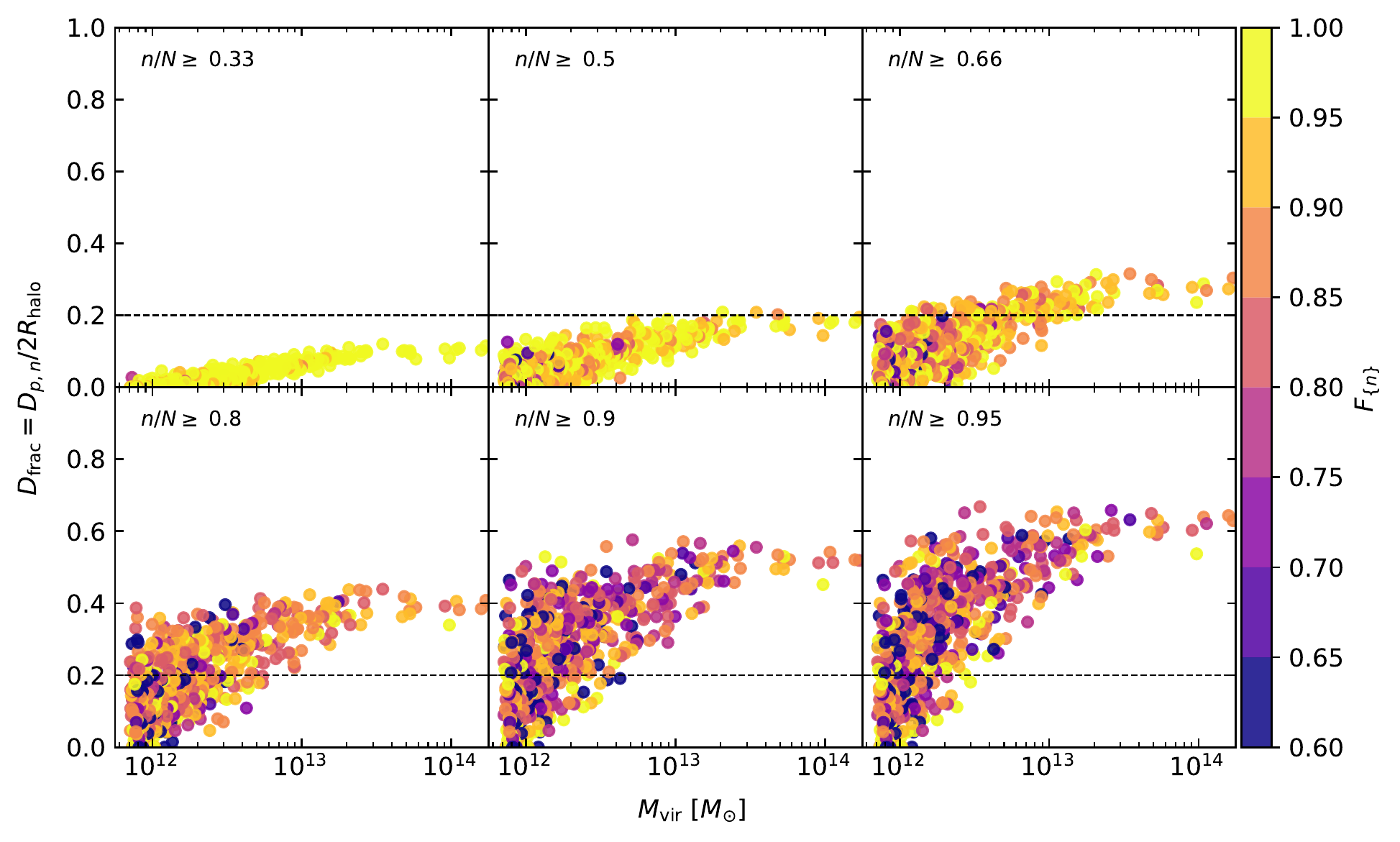}
    \caption{Thickness fraction $D_\mathrm{frac}$ of the thinnest planes of satellites satisfying different minimum satellite fractions $n/N$ as a function of the virial mass of the halo $M_\mathrm{vir}$. The horizontal dashed line marks our maximum thickness criterion $D_\mathrm{frac} \leq 0.2$. The colour-coding shows the in-plane momentum fraction $F_{\{n\}}$.}
    \label{fig:MTP_Complete_Multiplot_DOverM}
\end{figure*}

\section{Satellite Planes and Galaxy Properties}\label{sec:galprop}
With the method described above, we now test in how far the properties of the central galaxy are connected to the existence of a potential plane of satellites. The latter depends on the choice of thresholds for the three parameters that define the plane, namely
\begin{enumerate}
    \item The fraction $n/N$ of satellite galaxies $N$ included within the plane;
    \item The fraction $D_\mathrm{frac}$ of the maximum thickness of the plane with respect to the virial radius of the halo;
    \item The in-plane momentum fraction $F_{\{n\}}$, i.e., the minimum amount of the satellites' momentum that lies in the direction of the plane.
\end{enumerate}
In general, larger amounts of satellite galaxies $n/N$ and smaller values of $D_\mathrm{frac}$ lead to more plane-like satellite systems. And while satellites might also by chance simply appear to be in a plane at a given point in time, a large in-plane momentum fraction inside the plane indicates that the plane is indeed physically motivated and rotates as a plane around the galaxy, while a low in-plane momentum fraction suggests that the satellites might simply appear in a plane at some given point in time.

\subsection{Mass and Morphology}\label{subsec:mass_and_morphology}
For any given galaxy, its most defining features are its stellar mass, its total mass, and its morphology.
All three are intricately correlated to the galaxy's evolution pathway, and as such it is a very interesting question whether the existence of a satellite plane around a galaxy is correlated to these properties.

\subsubsection{Satellite planes with halo mass}
First, we investigate the planes formed by satellite galaxies as a function of the virial mass $M_\mathrm{vir}$ of the haloes, as the halo mass is directly correlated to the accretion history of the galaxy. Figure~\ref{fig:MTP_Complete_Multiplot_DOverM} shows the thickness fraction $D_\mathrm{frac}$ against virial mass $M_\mathrm{vir}$ for increasing satellite fractions $n/N \geq 0.33, 0.5$ and $0.66$ in the top row, and $n/N \geq 0.8, 0.9$ and $0.95$ in the bottom row. The in-plane momentum fraction $F_{\{n\}}$ is marked by the colour.

As can clearly be seen from Figure~\ref{fig:MTP_Complete_Multiplot_DOverM}, there is an overall trend with the virial mass for galaxies in more massive haloes to have on average thicker planes, independent of the satellite fraction that is included in the plane. Interestingly, we find that at the high mass end of groups and clusters there are planes of satellites even for fractions of 90\% of satellites being included that have a thickness of below 50\% and an in-plane momentum larger than 85\%. Further, if only 80\% of the satellites are included, the plane thickness in groups and clusters even decreases below 40\% for all such structures. This clearly demonstrates that planes of satellites do not only occur, but are even to be expected in the large bound structures of group and cluster environments, and not just around galaxies. This strongly indicates that there is a self-similarity from galaxy to galaxy cluster scales in terms of the existence of plane structures. 
This is in good agreement with the recent results from observations and from the dark matter-only \textsc{Millennium} simulation by \citet{gu:2022}.
Combined with the fact that the in-plane momentum is rather large even for these planes in clusters, this indicates that the accretion of satellites along filaments could be responsible for the occurrence of these planes, and that they are not just coincidental conglomerations. 

Furthermore, we see that planes with thickness below 20\% are a common occurrence in haloes of all virial masses if only half of the satellite galaxies are required to form the plane, as is evident from the upper central panel of Figure~\ref{fig:MTP_Complete_Multiplot_DOverM}. Under these conditions, the planes often have in-plane momenta that are larger than 90\%.
Figure~\ref{fig:MTP_Complete_Multiplot_DOverM} also highlights that when considering an increasing fraction of satellite galaxies in the planes, haloes with larger mass exhibit a larger thickness of the according plane. We interpret this as a result of the different formation mechanism of these haloes, where larger mass haloes grow predominately through major merger events \citep[e.g.,][]{oleary:2021}, which then naturally results in satellites to be distributed in thicker planes.
While there is an overall trend for the in-plane momentum to be larger with more satellite galaxies being included in the plane, there is no trend with virial mass for the in-plane momentum, as opposed to the plane thickness.

\begin{figure}
    \centering
    \includegraphics[width=\columnwidth]{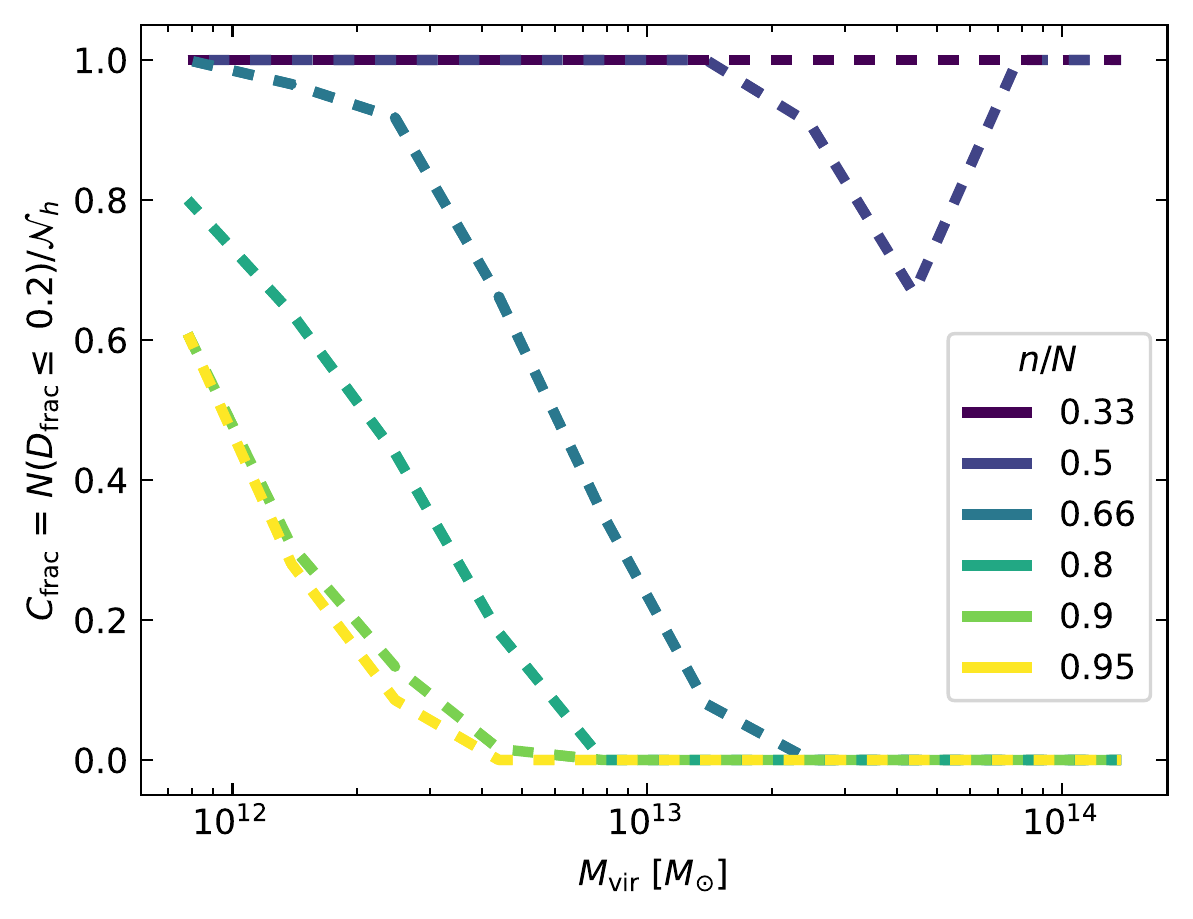}
    \caption{Fraction of satellite planes $C_\mathrm{frac}$  at a given virial mass $M_\mathrm{vir}$ fulfilling the criterion for the maximum thickness fraction $D_\mathrm{frac} \leq 0.2$ with increasing minimum satellites fractions $n/N$ as indicated in the legend.
    }
    \label{fig:MTP_Complete_Multiplot_DCriterionHistogram}
\end{figure}

The dashed lines in Figure~\ref{fig:MTP_Complete_Multiplot_DOverM} indicate a thickness of $D_\mathrm{frac} \leq 0.2$.
This maximum plane thickness level of 20\%  is used to more directly obtain the relative fractions of galaxies that have a plane with a thickness comparable to those found by \citet{Heesters_2021_MATLAS}. We show this in Figure~\ref{fig:MTP_Complete_Multiplot_DCriterionHistogram}, where we depict the fraction of satellite planes with plane thickness fractions below 20\%, for different satellite number fractions $n/N$, at a given virial mass. 
The fraction of systems fulfilling our maximum thickness criterion at a given mass decreases as the number of satellites included in the plane increases, as expected. Lower-mass haloes host thin satellite planes to a greater degree. 
Thus, we conclude that planes of satellites where (nearly) all satellite galaxies are visible inside a plane are more common and generally thinner for galaxies in low mass haloes, but if only a subset of the galaxies is required to be part of the plane, such planes can be found in haloes of Milky Way halo mass up to galaxy clusters.

\subsubsection{Satellite planes with galaxy morphology and stellar mass}
\begin{figure*}
    \centering
    \includegraphics[width=\columnwidth]{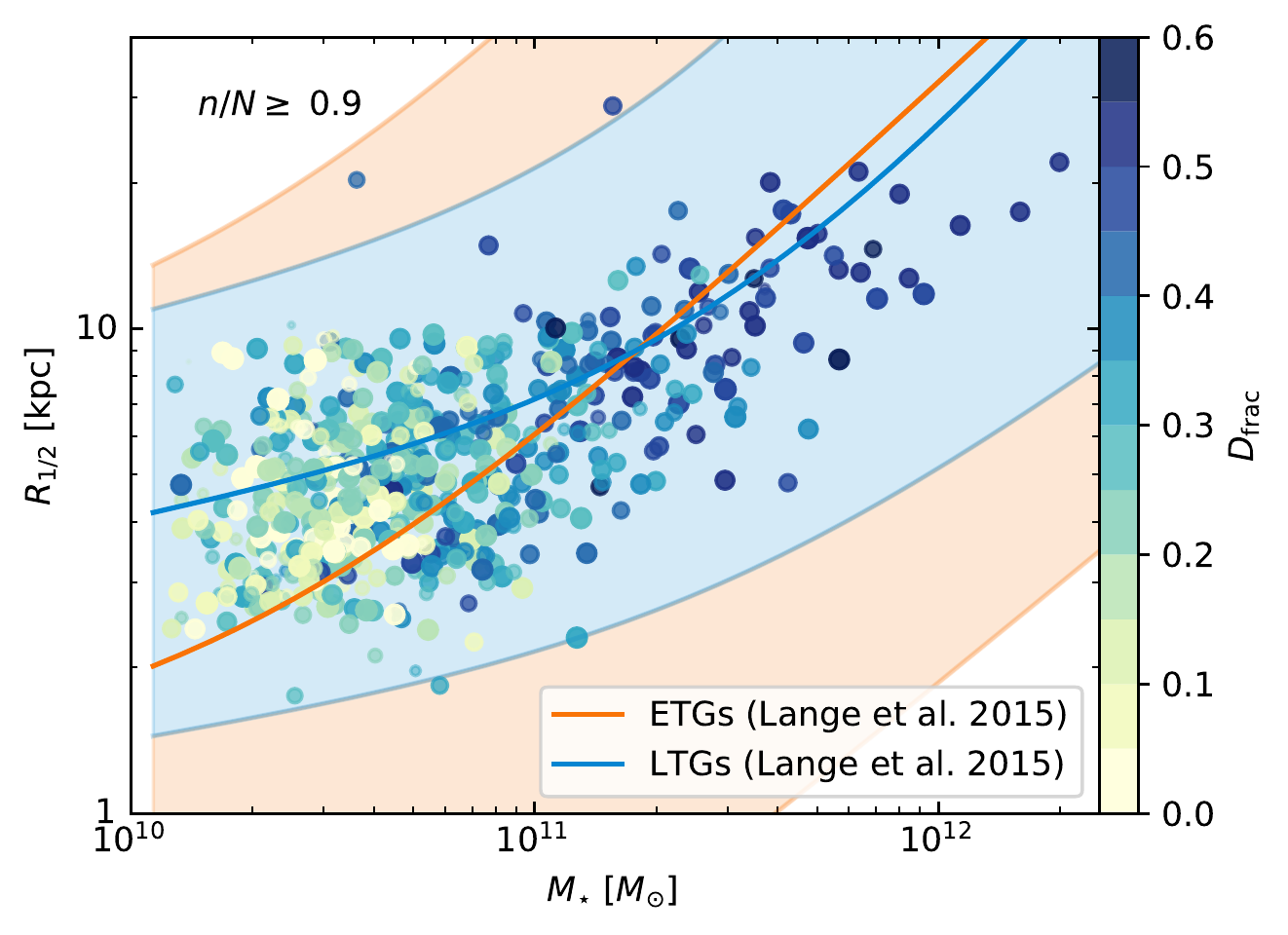}
    \includegraphics[width=\columnwidth]{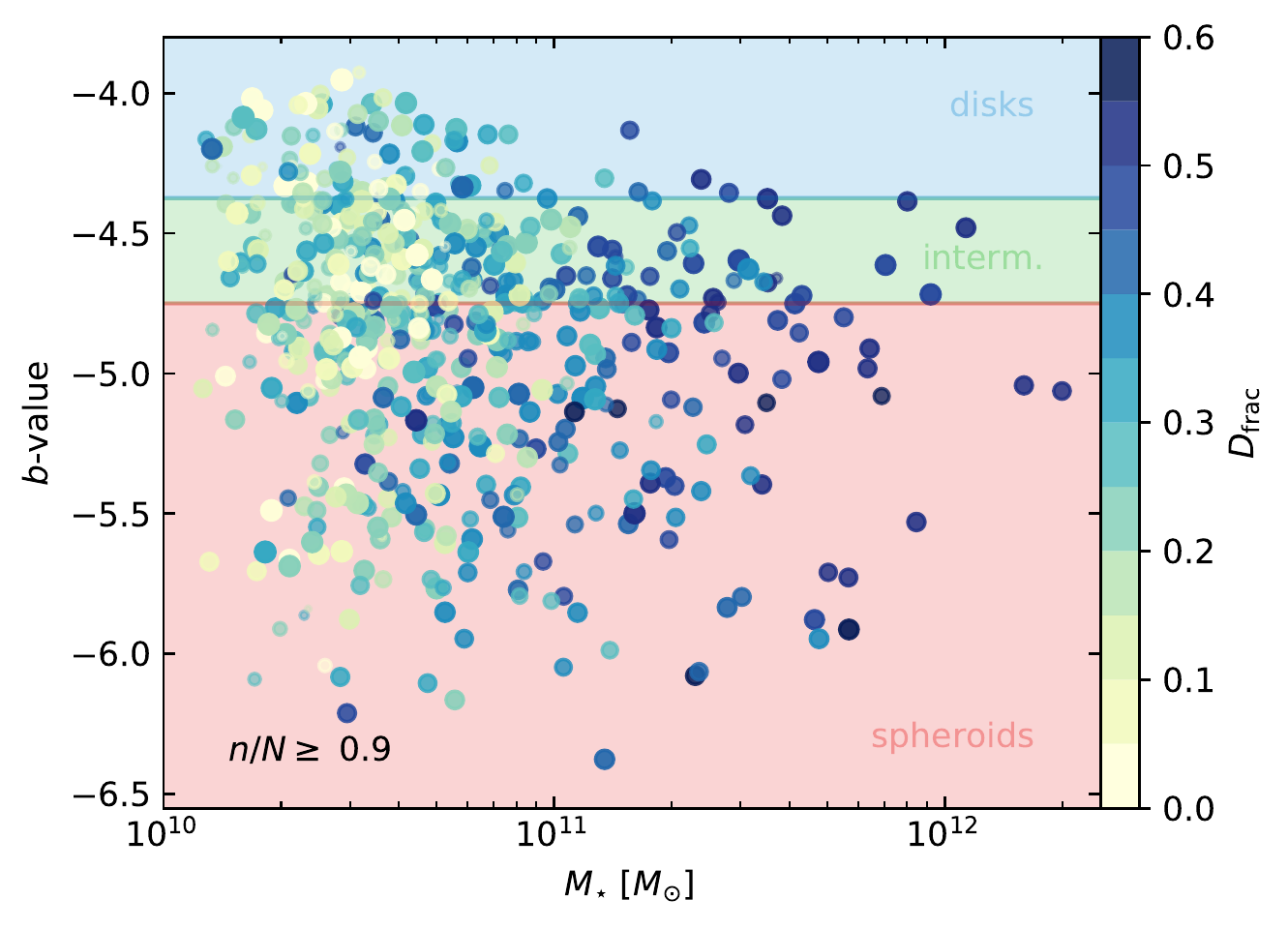}
    \includegraphics[width=\columnwidth]{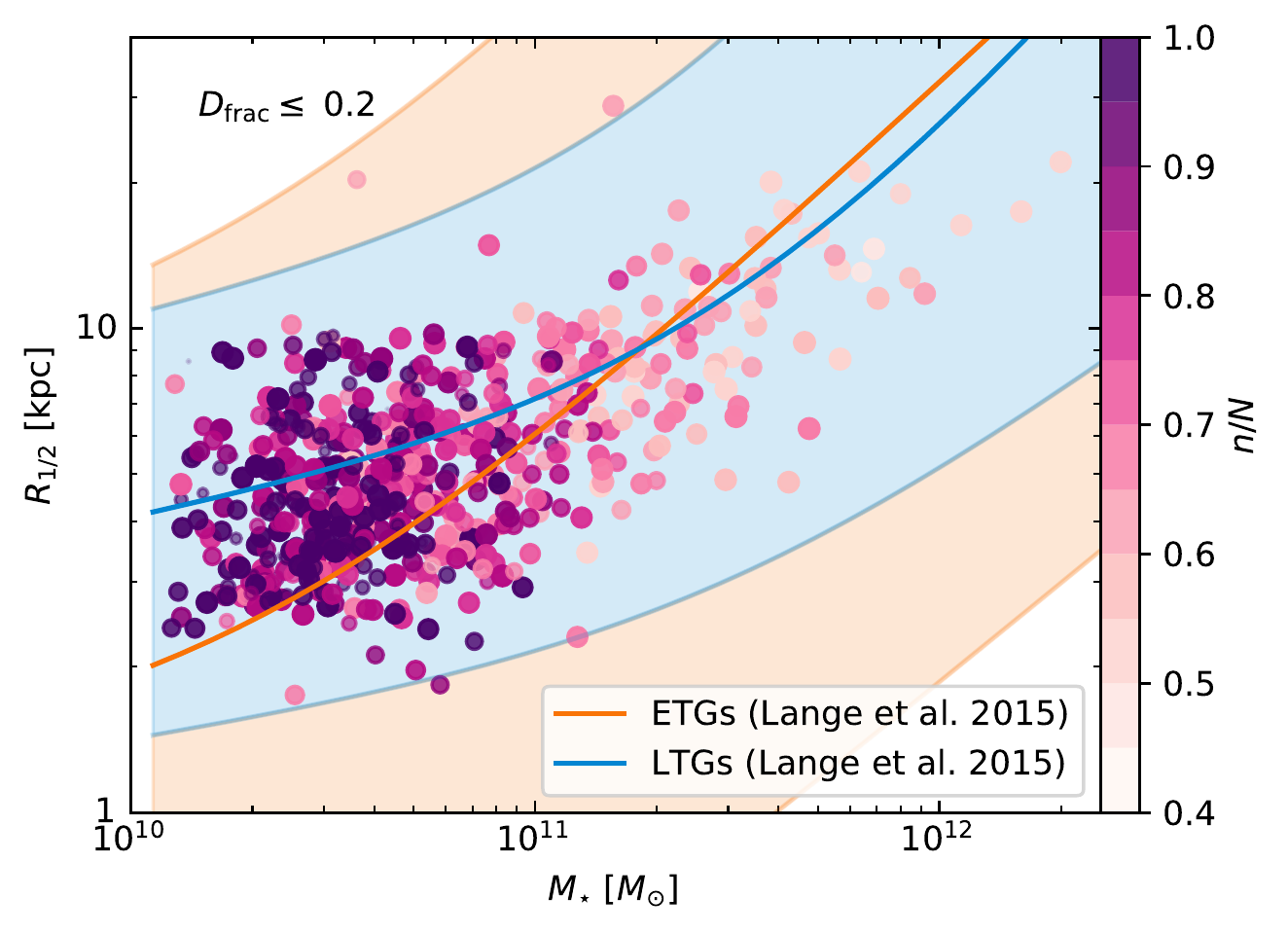}
    \includegraphics[width=\columnwidth]{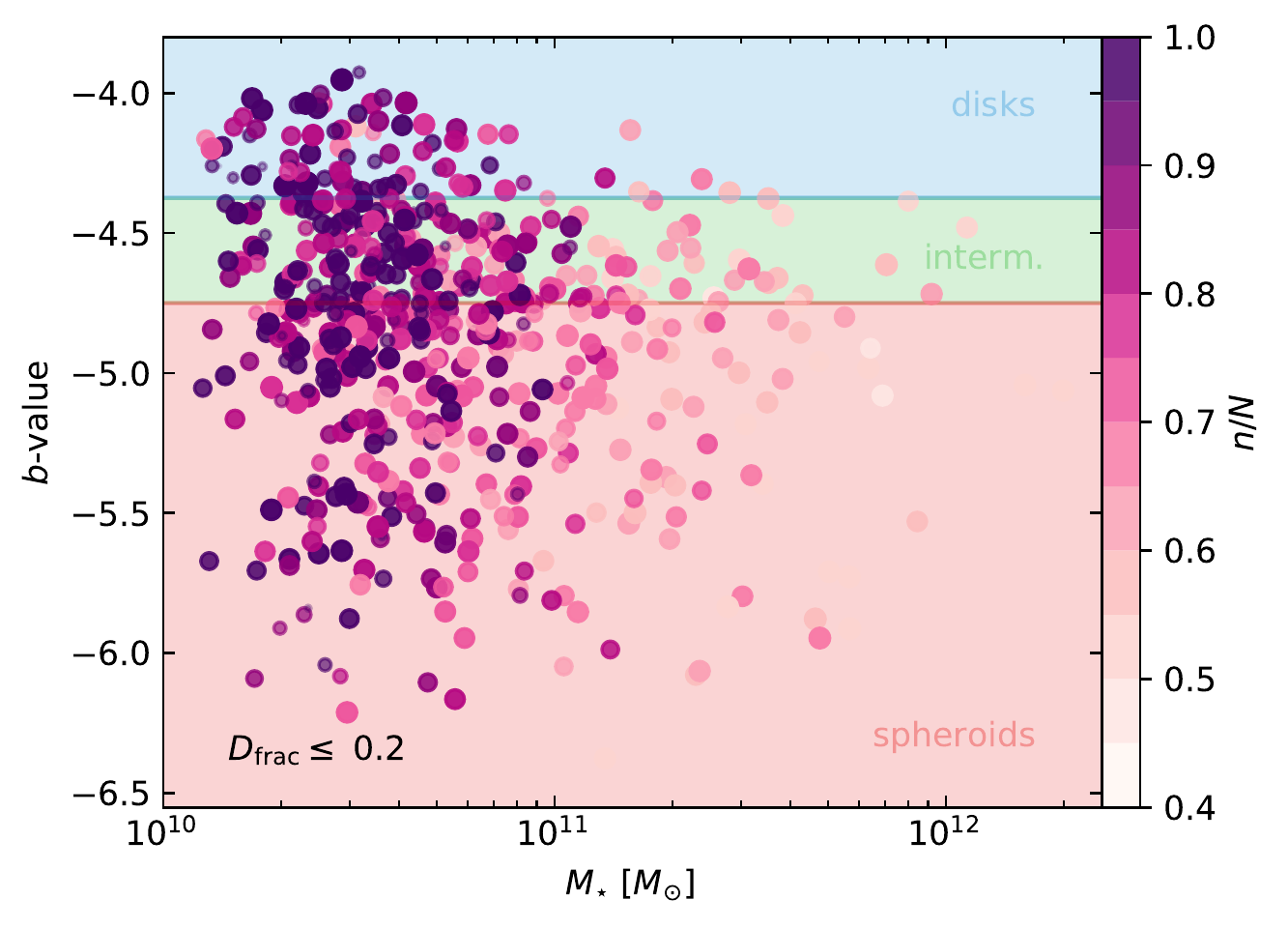}
    \includegraphics[width=\columnwidth]{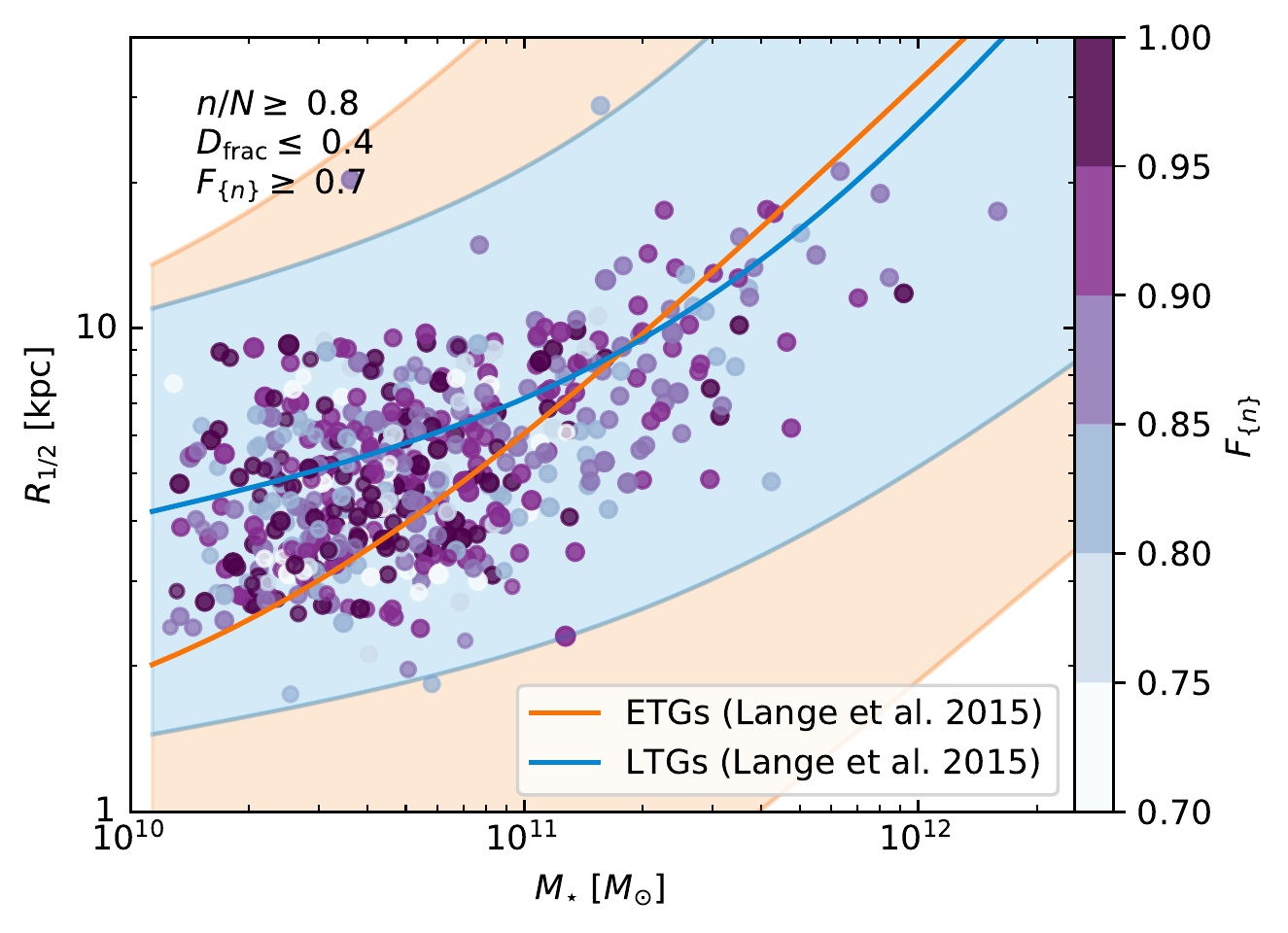}
    \includegraphics[width=\columnwidth]{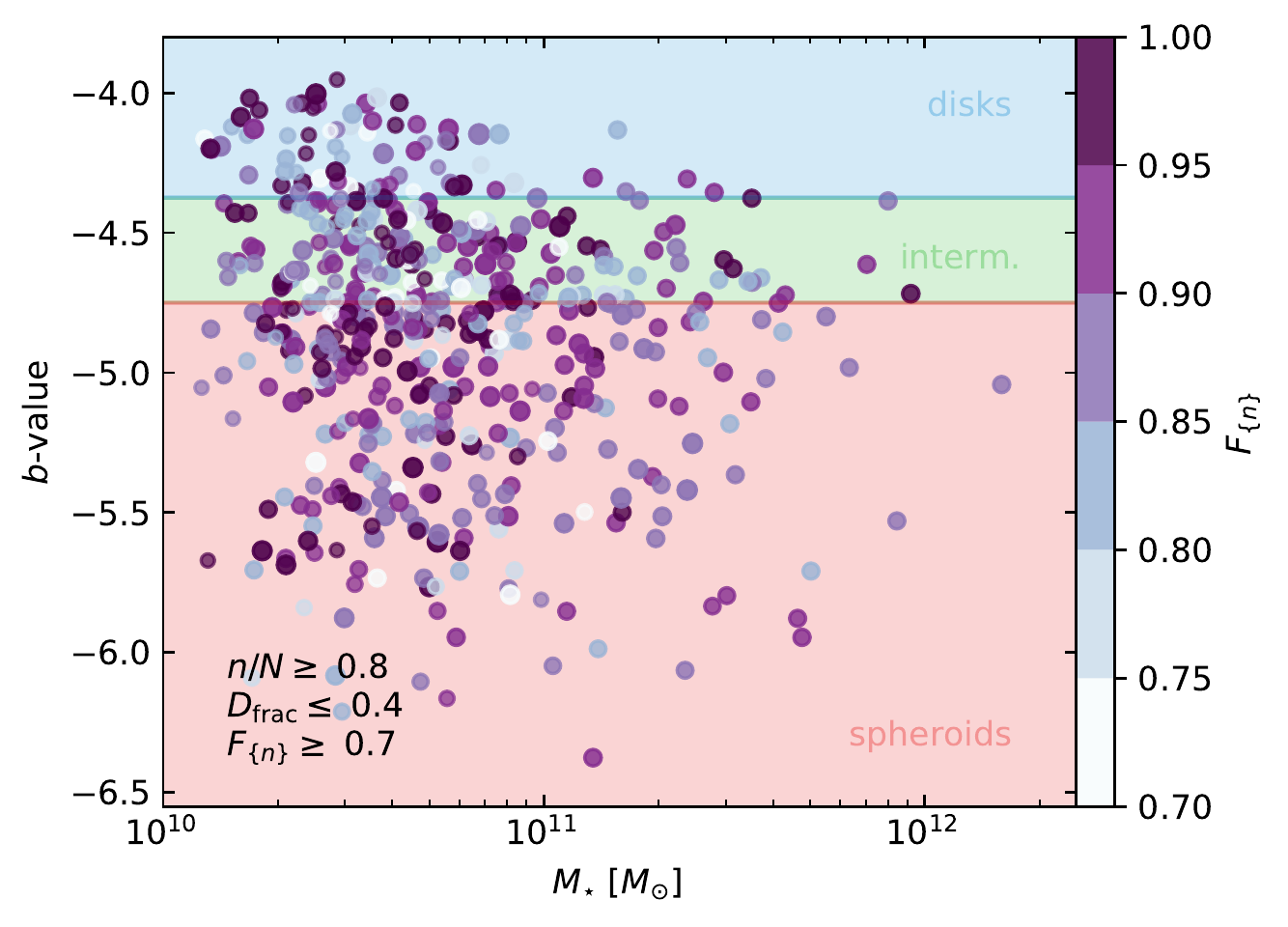}
    \caption{\textit{Left panels:}
    Mass-size relation ($M_*$ versus $R_{1/2}$) for the central galaxies in our sample, with observations by \citet{lange_2015_mass_size_relations} for the GAMA survey included as red and blue lines and shaded areas for ETGs and LTGs, respectively.
    \textit{Right panels:} Mass-morphology relation ($M_*$ versus $b$-value) for all central galaxies, with b-value cuts as given in Tab.~\ref{tab:b-values_in_our_data}.
    \textit{Top row:} For planes made of $n/N \geq 90\%$ of all satellite galaxies in the system. Colour according to the thickness fraction $D_\mathrm{frac}$, opacity and symbol size reflecting the in-plane momentum fraction $F_{\{n\}}$. 
    \textit{Middle row:} For planes with thickness of $D_\mathrm{frac} \leq 20\%$. Colour according to the satellite number fraction $n/N$, opacity and symbol size reflecting the in-plane momentum fraction $F_{\{n\}}$. 
    \textit{Bottom row:} For planes with $n/N \geq 80\%$, $D_\mathrm{frac} \leq 40\%$, and $F_{\{n\}} \geq 70\%$. Colour according to the in-plane momentum fraction $F_{\{n\}}$, opacity and size of the markers reflect the satellite number fraction $n/N$.
    }
    \label{fig:bval}
\end{figure*}
More intricately connected to a galaxy's formation history than its virial mass are its stellar mass and morphology. It is an established fact that galaxies occupy different areas in the mass-size plane, with elliptical galaxies being more compact than disk galaxies of the same stellar mass. This is the well observed mass-size relation \citep[e.g.,][]{lange_2015_mass_size_relations}. Another well know correlation that distinguishes disk galaxies from elliptical galaxies is their correlation between stellar mass and angular momentum, with disk galaxies of a given mass having larger angular momenta than ellipticals. The latter is known to be an excellent tracer for the morphology, finding its expression in the b-value as described by Equation~\ref{eq:bval} \citep{teklu_2015_angular_momentum_and_galactic_dynamics}.

The left-hand side panels of Figure~\ref{fig:bval} show the stellar mass versus half-mass radius for the central galaxies of our sample, in comparison to the observed mass-size relations for disk (late-type) galaxies (LTGs) and elliptical (early-type) galaxies (ETGs) from the GAMA survey \citep{lange_2015_mass_size_relations}. This relation has already been shown to be well reproduced by the galaxies from the \textsc{Magneticum} simulations by \citet{Schulze18,harris:20,Remus21} for the half-mass radii of the simulated galaxies calculated from the 3D distribution, and by \citet{Remus21} also for half-mass radii calculated from random 2D projections of the galaxies. 
We now show three different realisations of this relation, colouring the data points from the simulation according to different combinations of the three parameters $n/N$, $D_\mathrm{frac}$, and $F_{\{n\}}$ to define the existence of satellite planes.

In the upper left panel of Figure~\ref{fig:bval}, we require a plane to be assembled from at least 90\% of all satellites; the middle left panel depicts the results from the plane requirement of a maximum thickness of the plane of $D_\mathrm{frac} \leq 20\%$; the lowest panel requires a plane to contain at least 80\% of all satellites within a plane of thickness $D_\mathrm{frac} \leq 40\%$, as well as a minimum in-plane momentum of $F_{\{n\}} \geq 70\%$. As can clearly be seen, we do not find any trend with morphology, independent of the exact definition of the existence of a plane, and only a slight trend with stellar mass, similar to the tendencies already found for the virial mass: galaxies of larger central masses tend to have thicker planes than smaller mass galaxies for a fixed number of satellites being part of the plane. Similarly, if the thickness of the plane is fixed to thin planes, more massive central galaxies have less satellites from which this plane is build up. Interestingly, the trend with mass vanishes if the strict requirements are loosened and a plane is defined to contain ``only'' 80\% of the satellites with a plane thickness of up to 40\%, again demonstrating that these kind of planes are commonly found around galaxies of all types and masses.

The lack of any correlation of the existence of a plane with different morphological type is further supported by the right-hand side panels of Figure~\ref{fig:bval}, where we show the morphology directly against stellar mass, utilising the same three conditions as described before.
We confirm that independent of $b$-value, galaxies of lower stellar mass have a much higher prevalence of thinner, lower $D_\mathrm{frac}$ planes, and do not find a particular correlation between morphology and $D_\mathrm{frac}$. This clearly indicates that the existence of a satellite plane is independent from the current morphology of the central galaxy, which is in good agreement with the observations by \citet{crnojevic_2019_faint_end_cen_a_luminosity_function} and \citet{Heesters_2021_MATLAS}, and with the findings of Valenzuela et al.\ (in prep.), who found no correlation between the existence of satellite planes and the inner kinematics of galaxies in simulations and observations.
This result hints at the tidal dwarf scenario being an unlikely origin of the planes of satellites, as in such a case we would expect to see a prevalence of planes around elliptical galaxies, since they are much more prone to massive merger events than disk galaxies.
\begin{figure*}
\centering
    \includegraphics[width=\columnwidth]{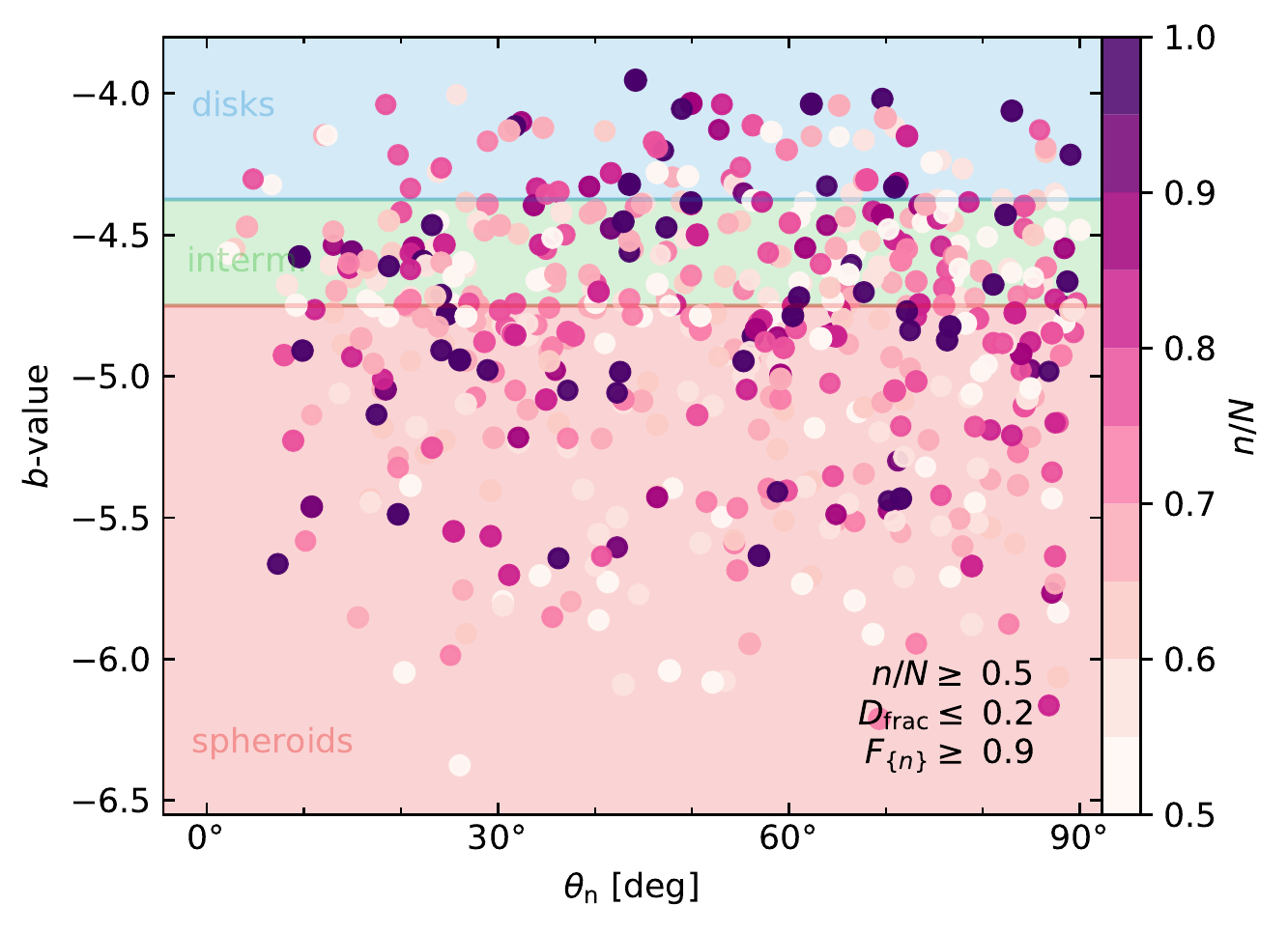}
    \includegraphics[width=0.95\columnwidth]{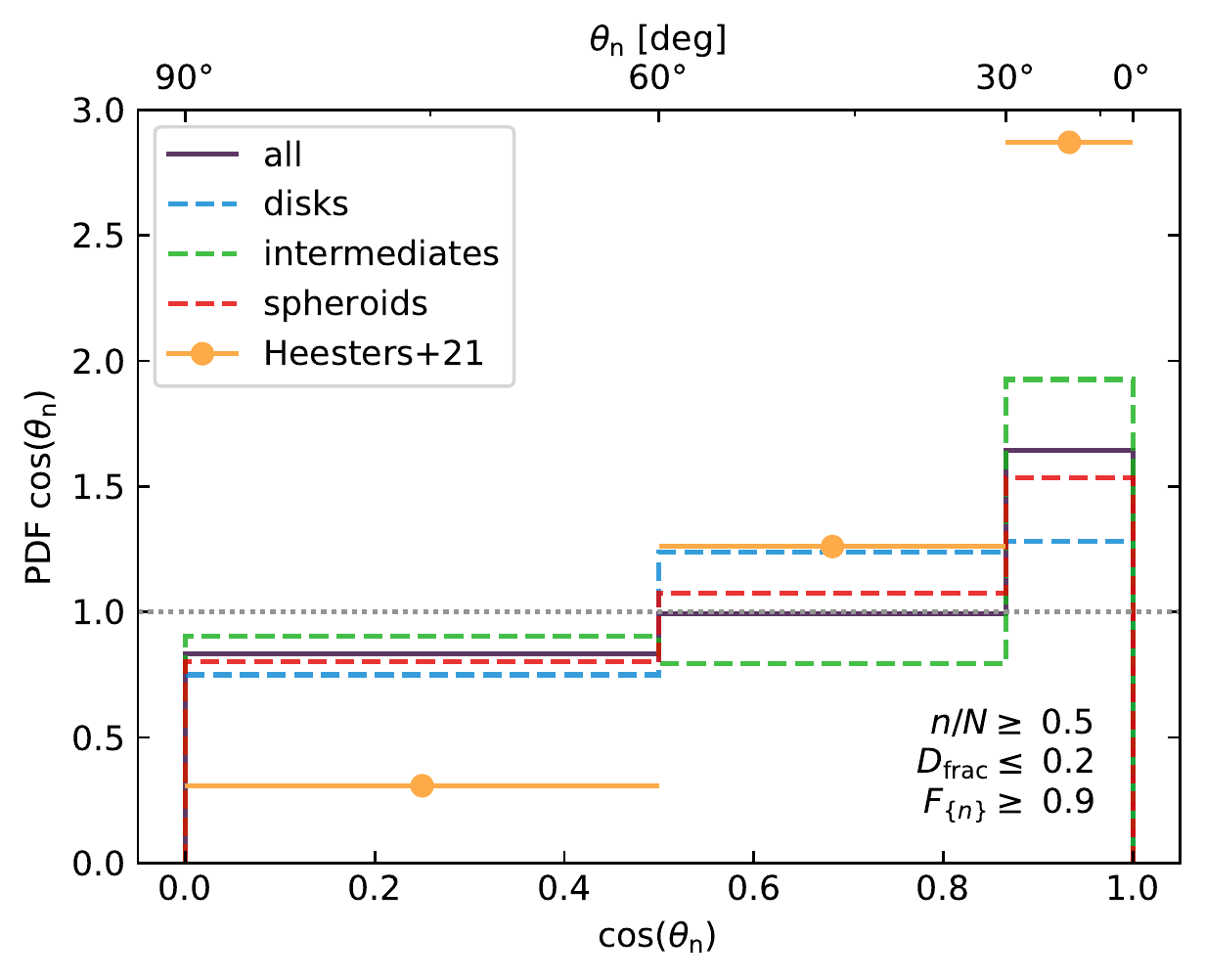}
    \caption{
    \textit{Left panel:} Polar angles $\theta_\mathrm{n}$ of the planes of satellite galaxies versus morphology as given by the $b$-value of the respective central galaxies, for planes consisting of at least $n/N \geq 50\%$ in a plane of at most $D_\mathrm{frac} \leq 20\%$ thickness and a minimum of $90\%$ in-plane momentum (see also top panel of Figure~\ref{fig:nplanes}). The colour reflects the satellite number fraction $n/N$ of each system's plane of satellites, while symbol size and opacity mark the respective in-plane momentum fraction $F_{\{n\}}$. 
    \textit{Right panel:} Probability density function of polar angles $\theta_\mathrm{n}$ of the planes of satellite galaxies for the same plane classification as in the left panel, with respect to $\cos(\theta_\mathrm{n})$. The distribution for all haloes fulfilling this criterion is shown in dark violet, while the distributions split according to $b$-value of the central galaxies are shown as dashed blue (disks), green (intermediates), and red (spheroids) histograms, respectively. 
    The orange circles with bars show the observational alignment measurements from \citet{Heesters_2021_MATLAS} after transformation into the uniform $\cos(\theta_\mathrm{n})$ parametrisation. 
    The grey dotted line marks the uniform random distribution of 3D vectors on a unit sphere, which is flat under a $\cos(\theta_\mathrm{n})$ parametrisation. 
    }
    \label{fig:bValuePlaneAngles}
\end{figure*}
\subsection{Alignment between the Satellite Plane and its Central Galaxy}
If the accretion of satellite galaxies along a plane is supposed to indicate the major direction of accretion infall onto a galaxy, and that direction is thought to be stable over time, we would generally expect to find an alignment between the minor axes of the satellite plane and the central galaxy. However, if no such alignment was found, that could either indicate that the direction of infall changes over time (or that a single group infall is the origin of the plane), and that the currently visible plane simply were different from the previous plane on which infall occurred. Alternatively, it could hint at that the planes are just aligned by chance at a given time, and their appearance has nothing to do with actual accretion events. 

We investigate the question of alignment in this section. To this end, we settle on the following definition for a satellite plane: we chose $n/N \geq 50\%$, $D_\mathrm{frac} \leq 20\%$, and $F_{\{n\}} \geq 90\%$, which basically means that the satellites are in a rather thin plane with a high in-plane momentum, but only a minimum of half of the satellites of the galaxies need to be part of that plane. We explore the impact of the minimum number of satellites and of the maximum thickness of the plane on the number of galaxies with detected planes in Appendix~\ref{app:nsat}.

By definition, the polar angle $\theta_\mathrm{n}$ already represents the alignment between the orientations of the galaxy's angular momentum and the plane normal vector, see Equation~\ref{eq:methods.angles}.
Figure~\ref{fig:bValuePlaneAngles} shows the angle $\theta_\mathrm{n}$ of all 535 galaxies with satellite planes that fulfil the above three-parameter criterion. 
The left-hand panel again shows the morphology as given by the $b$-value, compared to the polar angle $\theta_\mathrm{n}$, with the colour indicating the satellite number fraction $n/N$ of each system's plane of satellites and symbol size and opacity reflecting the respective in-plane momentum fraction $F_{\{n\}}$. The polar angle $\theta_\mathrm{n}$ is defined as the angle between the central galaxy's spin axis and the normal vector of the satellite plane. 
As can be seen immediately, we also do not find a correlation between the morphology of the galaxy and polar angle of the satellite plane, further supporting that the morphology of the central galaxy is completely independent of the current existence of a satellite plane. This is in disagreement with the results by \citet{welker_2017_caught_in_the_rhythm_2_comp_alignment,welker_2018_caught_in_the_rhythm_1_settle_into_plane} who found the alignment to be stronger for disks than for elliptical galaxies in the HorizonAGN simulations. In their work, however, the alignment is computed between all satellite positions relative to the central, as opposed to defining a thin plane which also has aligned velocities, as done in this work. This difference in methodology could be responsible for differences between the results from the two simulations.

While the left-hand panel of Figure~\ref{fig:bValuePlaneAngles} does not appear to show a strong preference for any particular orientation of the planes, we show in Appendix~\ref{app:uniform_random_on_sphere} that a randomly oriented plane in 3D space does not follow a uniform distribution of $\theta_\mathrm{n}$; instead, this is the case for $\cos(\theta_\mathrm{n})$. This means that small angles $\theta_\mathrm{n}$ are much less likely to occur for randomly oriented planes.

We adopt the transformation to $\cos(\theta_\mathrm{n})$ in the right-hand panel of Figure~\ref{fig:bValuePlaneAngles}, where we compute the probability density function (PDF) in three bins of $30$° each for all satellite planes fulfilling our three-parameter criterion (dark violet), as well as for central galaxies classified as disks (blue), intermediates (green) and spheroids (red). We also include observations from \citet{Heesters_2021_MATLAS} after transformation into the $\cos(\theta_\mathrm{n})$ parametrisation. The expectation for a uniform random distribution is shown as the horizontal grey dotted line.

We find a clear preference for satellite planes to be aligned with the spin of the central galaxy, in good agreement with observations: The excess probability for all planes in \textsc{Magneticum} in the well-aligned $\theta_\mathrm{n} = [0\degree, 30\degree]$ bin is $\xi_\mathrm{all} = 9\%$, with $\xi_\mathrm{disks} = 4\%$, $\xi_\mathrm{interm.} = 12\%$, and $\xi_\mathrm{spheroids} = 7\%$ for disk, intermediate and spheroidal centrals, respectively, and compute an excess probability of $\xi_\mathrm{obs} = 25\%$ in the observational measurements from \citet{Heesters_2021_MATLAS} over the same angle interval.
At the same time, we are in agreement with \citet{Heesters_2021_MATLAS} that a strong misalignment between the central galaxy and its satellite plane is disfavoured: in the close-to-perpendicular $\theta_\mathrm{n} = [60\degree, 90\degree]$ bin, we find negative excess probability of $\xi_\mathrm{all} = -8\%$ for all planes from the \textsc{Magneticum} simulations, with $\xi_\mathrm{disks} = -13\%$, $\xi_\mathrm{interm.} = -5\%$, and $\xi_\mathrm{spheroids} = -10\%$ for disks, intermediate and spheroidal centrals, respectively. The excess probability computed for that angle interval for the observational data from \citet{Heesters_2021_MATLAS} is found to be $\xi_\mathrm{obs} = -35\%$. While we cannot confirm the correlation between plane alignment and morphology reported by \citet{welker_2017_caught_in_the_rhythm_2_comp_alignment,welker_2018_caught_in_the_rhythm_1_settle_into_plane}, our results agree well in that alignment between galaxy and plane orientation is preferred.

We conclude that, while there is no correlation between the existence and orientation of a satellite plane and the morphology of the central galaxy, there is a clear tendency for satellite planes to be well aligned with the orientation of the central galaxy, independent of the morphological type. This clearly indicates that there is a strong likelihood for the planes to originate from continuous infall from the cosmic web with a primary infall direction.

\section{Summary and Conclusion}\label{sec:sumconc}
In this study we utilised more than 600 galaxies with masses above $M_\mathrm{vir} = 7.1 \times 10^{11} \, \mathrm{M_{\odot}}$ from the highest resolution volume of the fully hydrodynamical cosmological simulation suite \textsc{Magneticum Pathfinder} to investigate the existence of planes of satellites around such central galaxies.
To this end, we developed and introduced the \textit{Momentum in Thinnest Plane} (MTP) method to identify the planes of satellites around galaxies.

MTP determines and classifies satellite structures in a two-step process. First, by extensively sampling the space of spherical coordinates $(\theta,\phi)$ a broad range of potential plane orientations $p$ is created. Then, for every orientation the distances to the $n$ nearest satellites are determined, and only the one plane $p_\mathrm{n}$ is retained that is the thinnest such plane for a given number of satellites $n$. Finally, in a second step the momenta of all the constituting satellites are compared to the plane orientation to determine how well aligned the satellites' motions are to staying within the plane.

In using MTP, we have derived three parameters that describe the potential plane of satellites:
\begin{enumerate}
    \item $n/N$, the fraction of all satellite galaxies $N$ included within the plane.
    \item $D_\mathrm{frac}$, the fraction of the maximum thickness of the plane with respect to the virial radius of the halo.
    \item $F_{\{n\}}$, the in-plane momentum fraction, that is the amount of the satellites' momenta oriented parallel to the plane.
\end{enumerate}
Generally, we find that most of the planes that have small values of $D_\mathrm{frac}$, i.e., that are thin planes, also exhibit large values of the in-plane momentum $F_{\{n\}}$, in short: thin planes are usually constituted of satellite galaxies with large portions of their velocities parallel to the plane. This already indicates that the planes found by MTP are physical planes of satellites and not just coincidental alignments of galaxies at a snapshot in time.

Furthermore, we investigated the connection between the existence of a plane of satellites and properties of the central galaxy. More explicitly, we studied the virial masses, the stellar masses, the morphology as given by the $b$-value (the position of a galaxy in the angular momentum-stellar mass plane), and the angle between the minor axes of the plane and the central galaxy. The results are as follows:
\begin{itemize}
    \item Galaxies in more massive haloes (i.e., with larger virial masses) have on average thicker satellite planes, independent of the satellite fraction that is included in the plane. Even at the group and cluster mass range there are planes of satellites with 90\% of the satellites inside a plane of a thickness of below 50\% and an in-plane momentum larger than 85\%. Thus, \textit{planes of satellites are present even in group and cluster environments, not just around galaxies, highlighting the self-similarity from galaxy to galaxy cluster scales in terms of the existence of plane structures.} This result is in good agreement with observations of anisotropic planes in galaxy clusters that are thicker than their counterparts in galaxies but still present \citep{gu:2022}.
    \item There is only a slight trend with stellar mass: \textit{galaxies with smaller stellar masses tend to have thinner planes with generally more of their satellite galaxies inside such planes}, while galaxies of larger stellar masses tend to have either thicker planes or thin planes that consist of a smaller fraction of the total number of satellite galaxies, albeit that number is usually still larger than 50\% of all satellite galaxies.
    \item \textit{We do not find any correlation between the morphology of the central galaxy and the potential galaxy plane}, nor with the thickness of the plane, the in-plane momentum, or the number of satellite galaxies inside the plane. This disfavours the scenario in which planes are built up from tidal dwarf galaxies formed during merger events, as merger events are generally more prominent for elliptical galaxies and as such a correlation with morphology would be expected.
    \item We could also find \textit{no correlation between the morphology of the central galaxy and the polar angle of the plane}, in agreement with results from observations by \citet{Heesters_2021_MATLAS} and from simulations and observations by Valenzuela et al.\ (in prep.), but in disagreement with the results from the HorizonAGN simulations presented by \citet{welker_2017_caught_in_the_rhythm_2_comp_alignment,welker_2018_caught_in_the_rhythm_1_settle_into_plane}.
    \item Interestingly, we find a \textit{clear preference for the angular momenta of the galaxies and the satellite planes to be aligned, in agreement with the results from the MATLAS survey} \citep{Heesters_2021_MATLAS}. This indicates not only that the planes both in simulations and observations are actually physical in nature and not just coincidental conglomerations of satellites at a given point in time, supporting the idea that these planes of satellites originate from the infall of galaxies along the cosmic web and thus along a favoured direction with respect to the central galaxy.
\end{itemize}

We conclude that \textit{satellite planes around central galaxies from Milky Way masses up to galaxy cluster masses are common features, and are by no means in tension with $\Lambda$CDM}, supporting recent results presented by \citet{sawala:2022}. Furthermore, they are in fact \textit{indicators for the directional accretion of matter onto haloes along the cosmic web}, and could also be the cause of anisotropies found in the satellite systems of galaxy clusters \citep{2009A&A...501..419B,2013A&A...558A...1B}. The trend of planes being thicker with increasing host virial mass can be related to the increasing prevalence of major mergers for massive haloes as stated by \citet{oleary:2021}. However, using for example the Millenium and Millenium-II simulations, \citet{fakhouri:2008,fakhouri:2010} instead found a universal merger mass fraction for all haloes with only a very slight trend for higher mass haloes to have more massive mergers. In this case the here-found split in planes may instead be an imprint of the orientation and number of feeding filaments and thus of the cosmic web in general.
Thus, we expect thin satellite planes to be more commonly found in the future, owing to increasing detections of structures in the low surface brightness regime in observations. However, the question of the longevity of satellite planes remains to be answered.

\section*{Acknowledgements}
This work was supported by the Deutsche Forschungsgemeinschaft (DFG, German Research Foundation) under Germany’s Excellence Strategy - EXC-2094 - 390783311. KD, LCK and LMV acknowledge support by the COMPLEX project from the European Research Council (ERC) under the European Union’s Horizon 2020 research and innovation program grant agreement ERC-2019-AdG 882679. The calculations for the hydrodynamical simulations were carried out at the Leibniz Supercomputer Center (LRZ) under the project pr83li. We are especially grateful for the support by M.~Petkova through the Computational Center for Particle and Astrophysics (C2PAP). The hexagonal bins were done via \textsc{hexbin} in Matplotlib \citep{hunter07}.




\bibliographystyle{mnras}
\bibliography{satellite_planes_foerster_et_al_paper}



\appendix

\section{Uniform random distribution on a sphere}
\label{app:uniform_random_on_sphere}

\begin{figure}
  \centering
  \includegraphics{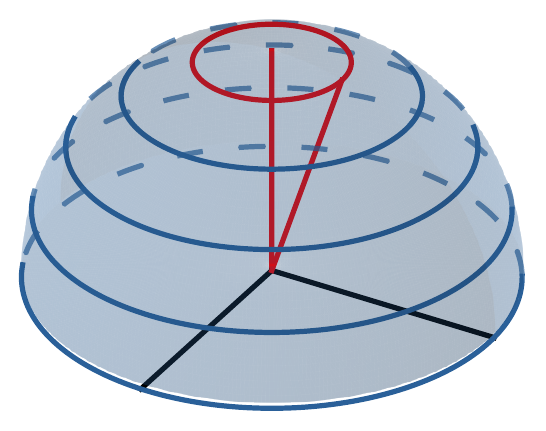}
  \caption{Visualisation of the derivation of the probability distribution function for randomly aligned axes, which is proportional to the circumference of a circle given by a polar angle, $\theta$. The angle between the two red lines is an example for such a polar angle.}
  \label{fig:sphere}
\end{figure}

For two randomly oriented axes, we find the probability distribution function (pdf) of the angle between them (in the range $0\degree$--$90\degree$) to be proportional to the circumference of a circle on a sphere given by the polar angle, $\theta$ (which corresponds to the alignment angle between the angular momentum vector and the normal of the plane). Such a polar angle is shown in Figure~\ref{fig:sphere} between the two red lines, for which the circle on the sphere is also highlighted in red. The circumference of a circle at a given polar angle, $\theta$, is $2\pi \sin\theta$, such that we obtain the pdf $p(\theta) \propto \sin\theta$. The proportionality factor is 1 since the pdf is already normalised:
\begin{equation}
    \int_0^{\frac{\pi}{2}} d\theta \sin\theta = 1,
\end{equation}
which means that the wanted pdf is \citep[also see e.g.,][]{ho&turner11}:
\begin{equation}
    p(\theta) = \sin\theta.
\end{equation}
The transformation $\theta \to y(\theta)$ that leads to a uniform distribution can be found through
\begin{equation}
    \int_0^{\theta'} d\theta \sin\theta = \int_{y(0)}^{y(\theta')} dy \frac{d\theta}{dy} \sin(\theta(y)),
\end{equation}
where the following equation must be fulfilled:
\begin{equation}
    \frac{d\theta}{dy} \sin(\theta(y)) \overset{!}{=} 1.
\end{equation}
This differential equation is solved by $\theta(y) = \arccos y$, which means that the uniform pdf for randomly oriented axes is given for $\cos\theta$ between 0 and~1.

\section{Number of satellite planes}\label{app:nsat}
\begin{figure}
    \centering
    \includegraphics[width=\columnwidth]{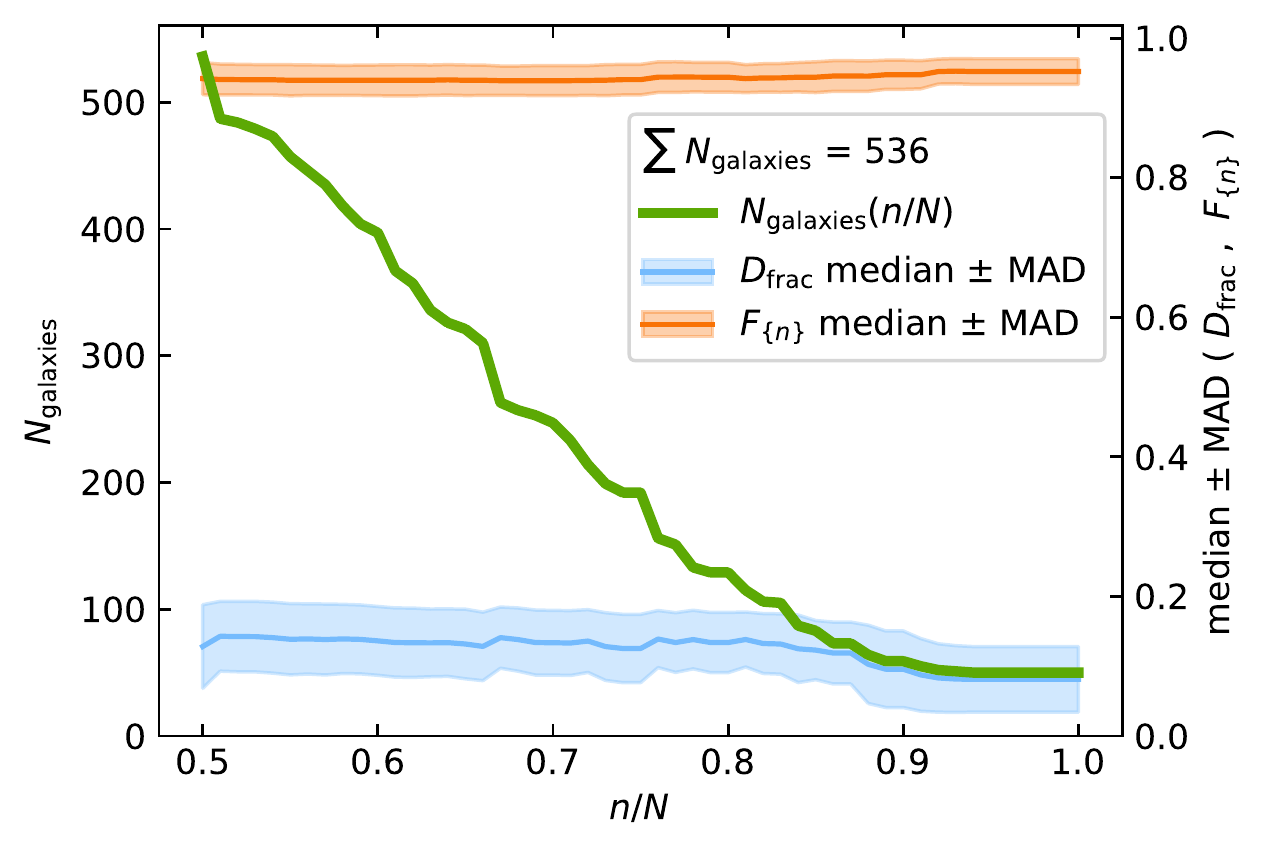}
    \includegraphics[width=\columnwidth]{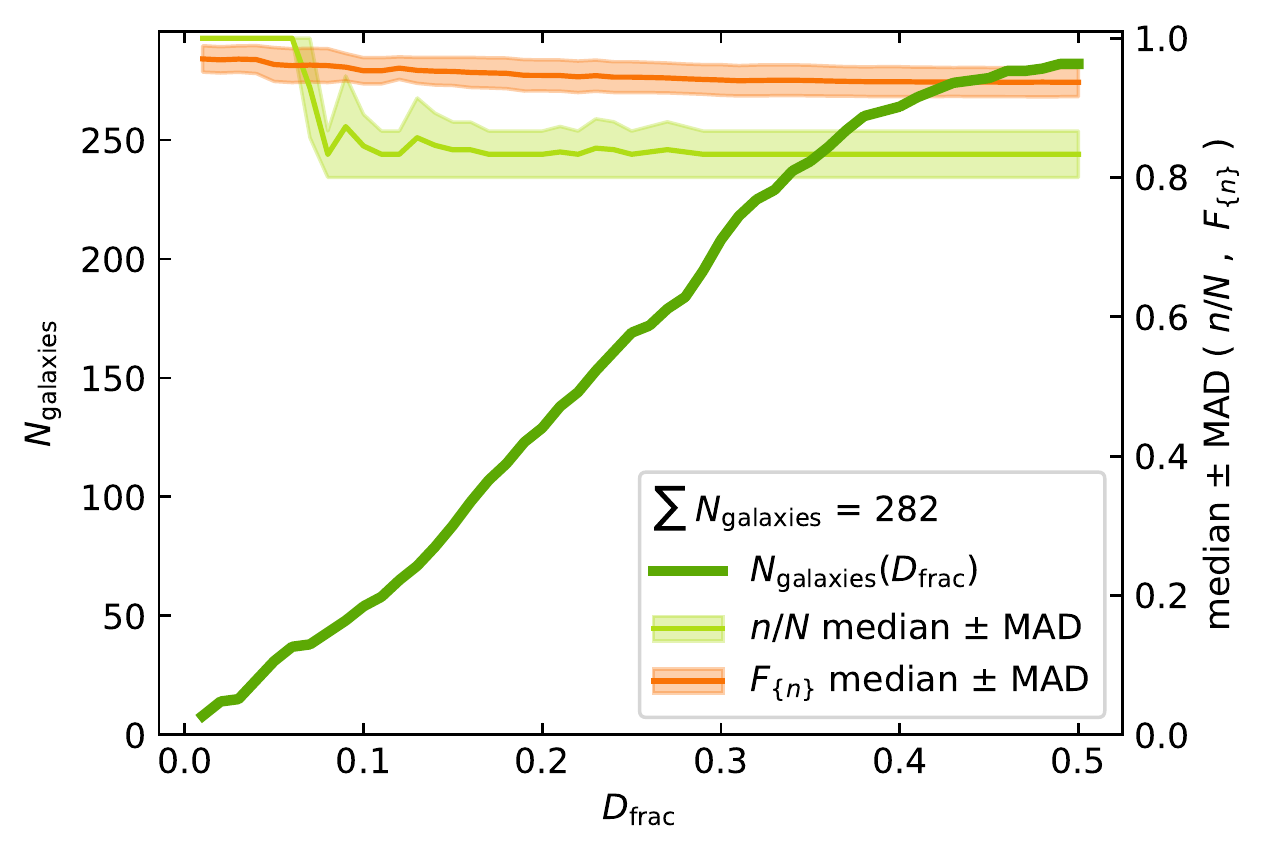}
    \caption{Number of galaxies with planes of satellite galaxies detected as a function of different quantities, depending on the criteria chosen. 
    \textit{Top panel:} $N_\mathrm{galaxies}$ as a function of the satellite number fraction $n/N$ (thick green line), as well as median and median absolute deviation of both thickness fraction $D_\mathrm{frac}$ (blue) and in-plane momentum fraction $F_{\{n\}}$ (orange). The criteria applied are $n/N \geq 0.5$, $D_\mathrm{frac} \leq 0.2$, and $F_{\{n\}} \geq 0.9$. The total number of galaxies satisfying these criteria is $\sum N_\mathrm{galaxies} = 535$. 
    \textit{Bottom panel:} $N_\mathrm{galaxies}$ as a function of the thickness fraction $D_\mathrm{frac}$ (thick green line), as well as median and median absolute deviation of both satellite number fraction $n/N$ (lime) and in-plane momentum fraction $F_{\{n\}}$ (orange). The criteria applied are $n/N \geq 0.8$, $D_\mathrm{frac} \leq 0.5$, and $F_{\{n\}} \geq 0.9$. The total number of galaxies satisfying these criteria is $\sum N_\mathrm{galaxies} = 281$. 
    }
    \label{fig:nplanes}
\end{figure}
We examine the number of galaxies for which satellite planes are detected with the MTP method, depending on either the required fraction of satellites in the plane, or on the maximum allowed thickness fraction of the plane. Figure~\ref{fig:nplanes} shows the number of galaxies with planes, $N_\mathrm{galaxies}$, as a function of the minimum satellite number fraction $n/N$ in the top panel, and as a function of the thickness fraction in the bottom panel. 
\begin{itemize}
    \item Top panel: we require a minimum of $n/N \geq 50\%$ of satellites to be part of the plane, the plane to be at most $D_\mathrm{frac} \leq 20\%$ thick, and the momentum of the satellites to be at least $F_{\{n\}} \geq 90\%$ in the direction of the plane. The thick green line shows the cumulative number of planes that consist of $n/N$ or more of the satellites, while the blue and orange lines and shaded regions represent the median and mean absolute deviation of $D_\mathrm{frac}$ and $F_{\{n\}}$ of the planes, respectively.
    We find a total number of $N_\mathrm{galaxies} = 535$ that simultaneously fulfil these criteria. We use this set of criteria to identify satellite planes in this work.
    \item Bottom panel: we require a minimum of $n/N \geq 80\%$ of satellites to be part of the plane, the plane to be at most $D_\mathrm{frac} \leq 50\%$ thick, and the momentum of the satellites to be at least $F_{\{n\}} \geq 90\%$ in the direction of the plane. The thick green line shows the cumulative number of planes that are $D_\mathrm{frac}$ or less in thickness fraction, while the lime and orange lines and shaded regions represent the median and mean absolute deviation of $n/N$ and $F_{\{n\}}$ of the planes, respectively.
    We find a total number of $N_\mathrm{galaxies} = 281$ that simultaneously fulfil these criteria, pointing to a population of galaxies with thicker, plane-like structures consisting of the vast majority of their respective satellites that move in the direction of the plane.
\end{itemize}


\bsp	
\label{lastpage}
\end{document}